\documentclass[prb, twocolumn, notitlepage, superscriptaddress, eqsecnum ]{revtex4-2}
\usepackage{amsmath,amsfonts, amssymb, amsthm, dsfont}
\usepackage{yfonts}
\usepackage{bm}
\usepackage{mathrsfs}
\usepackage{graphicx}
\usepackage{verbatim}
\usepackage{hyperref}
\usepackage{wasysym}
\usepackage{xcolor}
\usepackage{tikz}
\usepackage{physics}
\usepackage{subfigure}

\DeclareMathOperator*{\SumInt}{%
\mathchoice%
  {\ooalign{$\displaystyle\sum$\cr\hidewidth$\displaystyle\int$\hidewidth\cr}}
  {\ooalign{\raisebox{.14\height}{\scalebox{.7}{$\textstyle\sum$}}\cr\hidewidth$\textstyle\int$\hidewidth\cr}}
  {\ooalign{\raisebox{.2\height}{\scalebox{.6}{$\scriptstyle\sum$}}\cr$\scriptstyle\int$\cr}}
  {\ooalign{\raisebox{.2\height}{\scalebox{.6}{$\scriptstyle\sum$}}\cr$\scriptstyle\int$\cr}}
}
\newcommand{\tph}{\Tilde{\varphi}}
\newcommand{\tth}{\Tilde{\theta}}
\newcommand{\bphi}{\bar{\varphi}}
\newcommand{\bth}{\bar{\theta}}

\newcommand{\di}{\mathrm{d}}

\renewcommand{\vr}{{\mathbf{r}}}
\newcommand{\vp}{{\mathbf{p}}}

\newcommand{\comments}[1]{}
\newcommand{\mb}[1]{\mathbf{#1}}
\renewcommand{\cal}[1]{\mathcal{#1}}
\newcommand{\p}{\prime}

\def\U{\mathrm{U}(1)}

\def\Z{\mathbb{Z}}
\def\TT{\mathsf{T}}

\DeclareMathOperator{\sgn}{sgn}

\makeatletter
\def\l@subsubsection#1#2{}
\makeatother

\begin{document}

\title{Weak symmetry breaking and topological order in a 3D compressible quantum liquid}
\author{Joseph Sullivan}
\affiliation{Department of Physics, Yale University, New Haven, CT 06511-8499, USA}
\author{Arpit Dua}
\affiliation{Department of Physics and Institute for Quantum Information and Matter, California Institute of Technology, Pasadena, California 91125, USA}
\author{Meng Cheng}
\affiliation{Department of Physics, Yale University, New Haven, CT 06511-8499, USA}
\date{\today}

\begin{abstract}
   We introduce a new type of 3D compressible quantum phase, in which the U(1) charge conservation symmetry is weakly broken by a rigid string-like order parameter, and no local order parameter exists. We show that this gapless phase is completely stable and described at low energy by an infinite-component Chern-Simons-Maxwell theory. We determine the emergent symmetry group, which contains U(1) 0-form planar symmetries and an unusual subgroup of the dual U(1) 1-form symmetry supported on cylindrical surfaces. Through the associated 't Hooft anomaly, we examine how the filling condition is fulfilled in the low-energy theory. We also demonstrate that the phase exhibits a kind of fractonic topological order, signified by extensively many different types of topologically nontrivial quasiparticles formed out of vortices of the weak superfluid. A microscopic model realizing the weak superfluid phase is constructed using an array of strongly coupled Luttinger liquid wires, and the connection to the field theory is established through boson-vortex duality.
\end{abstract}

\maketitle

\tableofcontents

\section{Introduction}

A quantum state is compressible if the charge density can be tuned continuously by changing the chemical potential in a translation-invariant system. 
Common examples of stable compressible matter include:
\begin{enumerate}
    \item Symmetry-breaking phase, in particular when the U(1) charge conservation is spontaneously broken by a local order parameter, i.e. a superfluid.
    \item For fermionic systems, the Fermi liquid is a compressible phase which does not break any symmetry (U(1) or translation). 
\end{enumerate}
Both can be ``deformed" to obtain other compressible phases with distinct low-energy dynamics. For example, a Fermi liquid can be coupled to gapless bosonic excitations (e.g. critical fluctuations or gauge fields), leading to various kinds of non-Fermi liquids. For a superfluid, by adding additional interactions one can change the dispersion of the Goldstone boson to alter the low-energy physics. Examples of this include the quantum Lifshitz liquid~\cite{Ardonne_2004} and Bose-Luttinger liquid~\cite{SurPRB2019, lake2021boseluttinger}.

 While it is a formidable task to generally classify the low-energy dynamics of gapless phases, one can first characterize them at the kinematic level, in particular their emergent symmetries and the associated quantum anomalies. 
Recently Ref. [\onlinecite{ElsePRX2021}] has taken this approach to study the low-energy theory of compressible phases.  The filling condition can be formulated in a way similar to the 't Hooft anomaly~\cite{ChengPRX2016, ChoPRB2017}, and  in particular it is shown that the emergent symmetry group cannot be a compact 0-form and/or finite higher-form symmetry group. The previous two classes of examples illustrate two mechanisms to satisfy the filling condition: the symmetry group of the superfluid phase is $\U^{[0]}\times\U^{[D-1]}$~\cite{Delacretaz_2020, ElseDrag2021}, where $D$ is the spatial dimension. The presence of the $\U^{[D-1]}$ form symmetry is essentially equivalent to the fact that superfluid vortices must form $(D-1)$-dimensional closed surface. The mixed anomaly between the two subgroups is responsible for the filling condition. The same mechanism underlies some more exotic examples of compressible phases of bosons, such as the Bose-Luttinger liquid~\cite{lake2021boseluttinger}. For the Fermi liquid, the emergent symmetry is the (0-form) loop group\cite{ElsePRX2021} $\mathrm{L}^{D-1}\U$, which  describes at the level of kinematics the existence of a Fermi surface. The group is ``larger" than any compact Lie group.  Deformations of the two types of examples may change the emergent symmetry group, but do not affect the way that the filling anomaly is realized. In a sense, these two examples represent two ``minimal" emergent symmetry groups that allow compressibility in the respective classes. It is clearly an important problem to understand whether there exist other low-energy kinematical structures compatible with a generic filling factor. 

Ref. [\onlinecite{ElsePRX2021}] mostly operates within the framework of quantum field theory in continuum with both translation and rotation symmetries. The aforementioned examples can be easily modified to introduce certain anisotropies, for example in boson velocity or the shape of the Fermi surface. However these modifications can all be smoothly turned off without changing the nature of the states i.e. no phase transitions. In particular, they do not affect the kinematical structure.  A natural question then is whether there are fundamentally new types of symmetry mechanisms to allow arbitrary filling, once we go beyond the framework of continuum field theory. In fact, there is a trivial toy example that goes beyond what we have discussed so far: consider a stack of decoupled 2D superfluids~\footnote{Similarly, one can consider a stack of 2D Fermi liquids. However, such a state has an open Fermi surface consisting of two curves wrapping around the Brillouin zone.}. Such a phase is obviously compressible but the emergent symmetry group is very different from that of a 3D superfluid. More specifically, each layer has a separate $\U$ 0-form symmetry and $\U$ 1-form symmetry at low energy. One may object that a stack of 2D superfluids is unstable to infinitesimal boson tunneling, thus does not represent a truly stable phase. Quantum phases with such ``subsystem" symmetries have been vigorously studied in recent years, largely inspired by the discovery of fracton topological order~\cite{Chamon2005, BravyiAOP2011, Haah, VijayPRB2015, VijayPRB2016, WilliamsonPRB2016, Nandkishore2018review, Pretko2020review}. It has been recognized recently that a large class of two- and three-dimensional quantum phases cannot be described as conventional quantum field theories~\cite{XuPRB2006, Rasmussen2016, PretkoPRB2017a, PretkoPRB2017b,  PretkoGauge2018, Seiberg_2021a, Seiberg2020, Seiberg_2021b,  SlaglePRB2017, SlaglePRL2021}. They are often characterized by exotic global symmetries~\cite{WilliamsonPRB2016, Shirley_2019, you2018subsystem}, e.g. those defined on certain submanifolds.  

Motivated by this question and building on top of recent works on fracton models~\cite{ma2020fractonic, FractonCW2021, JoeCW2021}, in this work we study a new example of a completely stable compressible state in three dimensions, whose low-energy sector is described by an infinite-component Chern-Simons (iCS) theory, consisting of coupled planar U(1) gauge fields (other generalizations of CS-like theory to 3D in the context of fractonic order have been considered in [\onlinecite{YouCS}] and [\onlinecite{WilliamsonPRB2019}]). Variants of the field theory with fully gapped spectrum have been studied in Ref. [\onlinecite{ma2020fractonic}], which are shown to possess  a particular kind of fracton topological order, with all quasiparticles restricted to move in planes. Moreover, it is proposed that the theory provides an example of a ``non-foliated" fracton topological phase. Ref. [\onlinecite{ma2020fractonic}] also noticed that the field theory we obtain has a gapless spectrum. One motivation for this work is to elucidate the rich physics of the gapless iCS theory, and provide a lattice realization which can be controllably solved and which manifests the symmetries of the underlying field theory.

Below we briefly summarize the main features of this compressible quantum liquid, focusing on those that distinguish it from the known types of compressible phases previously mentioned. Most importantly, the emergent symmetry group (relevant for the filling anomaly) consists of a 0-form $\U$ symmetry group of charge conservation, and a ``cylindrical" 1-form symmetry, the meaning of which will be defined later. (Note that an ordinary 3D superfluid has an emergent U(1) 2-form symmetry.) This exotic emergent 1-form symmetry stems from a rigid string-like superfluid order, where an order parameter for the U(1) 0-form symmetry is supported on a straight line penetrating the entire system. However, there exists no other local order parameter. Thus we dub this phenomenon ``weak symmetry breaking" (WSB). The terminology has been previously used in the context of (2+1)d symmetry-enriched topological order~\cite{Kitaev_2006, WangPRB2013, Rao_2021}, when there exists no local order parameter but certain non-local observables break the (0-form) symmetry. In addition, the low-energy theory also exhibits finite 1-form symmetries, corresponding to deconfined, topologically nontrivial quasiparticle excitations. They can be formed from dipoles of vortices of the ``weak" superfluid. Thus in a sense the phase of matter intertwines U(1) symmetry breaking and topological order in an interesting way.

Phenomenologically,  we find that at low energy there is only a single gapless point. We argue that the gapless state is in fact robust to any local perturbation and therefore represents a stable gapless phase. In particular, the gapless excitations are charge neutral and all local charged excitations are gapped. We also study the transport properties, and find that the phase exhibits a superconducting response in the plane perpendicular to the direction of the non-local order parameter, and insulating in the other.

The paper is organized as follows: in Sec. \ref{sec:Coupled wire construction} we first provide a microscopic realization of the compressible liquid with WSB, using a coupled-wire construction, and study its various properties. We derive an infinite-component CS theory description of the low-energy physics, by applying boson-vortex duality transformation in the coupled wire setting. Gapped cases of these field theories have been recently studied in Ref. [\onlinecite{ma2020fractonic}] as examples of Abelian fractonic phases without any foliation structure. In Sec. \ref{sec:Layered CS gauge theory} we then explore the emergent symmetries, anomalies (especially the filling anomaly) and stability of the phase, using the dual gauge theory formulation.

\section{Coupled wire construction}
\label{sec:Coupled wire construction}
We start from a microscopic model realizing the phase. The model is built from an array of interacting one-dimensional quantum wires. Such models have proven very fruitful in providing explicit realizations of topological phases in both two and three dimensions \cite{KanePRL2003, TeoKane2014, FractonCW2021, JoeCW2021, fuji2019CW, JoeCW2021, FractonCW2021}.

\subsection{Model Hamiltonian and symmetries}
\label{sec:Model Hamiltonian and symmetries}
Consider quantum wires arranged in a square lattice. Each wire is described by a $c=1$ bosonic Luttinger liquid, with a K matrix $K_\text{w}=\sigma^x$. We postulate that the Luttinger liquid is realized as the low-energy effective theory of a one-dimensional chain of bosons (or spins). The Hamiltonian is
\begin{equation}
\label{eq:free boson H}
	H=\frac{v}{2\pi}\sum_\mb{r}\int\di x\, [(\partial_x\varphi_\mb{r})^2+(\partial_x\theta_\mb{r})^2].
\end{equation}
 Here $\vr=(y,z)$ labels the position of wires in the $yz$ plane. The bosonic fields $\varphi$ and $\theta$ satisfy the canonical commutation relation
\begin{equation}
	[\varphi_\mb{r}(x_1), \partial_{x_2}\theta_{\mb{r}'}(x_2)]=2\pi i \delta(x_1-x_2)\delta_{\mb{r}\mb{r}'}.
	\label{}
\end{equation}
We have also assumed that all wires have the same velocity to ensure translation invariance.

\begin{figure}[t!]
    \centering
    \includegraphics[width=\columnwidth]{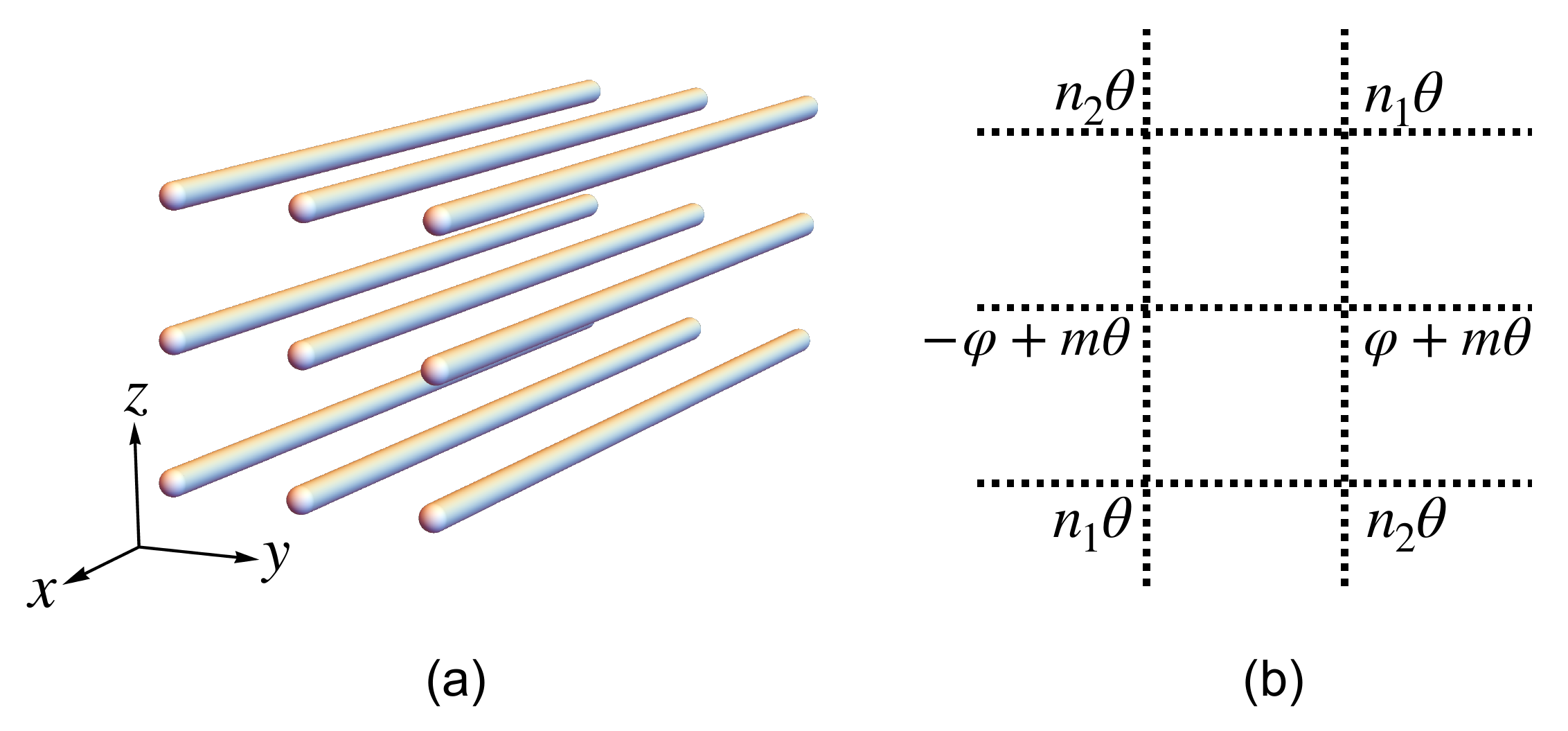}
    \caption{The coupled wire construction in this work starts from a 2D array of single-component Luttinger liquids, as illustrated in (a), and then gapping interactions between nearby wires are turned on to create a 3D quantum state. The form of the interaction is illustrated in (b).}
    \label{fig:couple-wire}
\end{figure}

We add the following type of interactions to gap out the wires:
\begin{equation}
	-g\sum_{\mb{r}}\int\di x\, \cos \Theta_{\mb{r}}(x), g>0.
	\label{}
\end{equation}
Here $\Theta_{\mb{r}}\equiv\Theta_{y,z}$ is defined as 
\begin{equation}
\begin{split}
	\Theta_{y,z}&= -\varphi_{yz}+ m \theta_{yz} + \varphi_{y+1,z}+m\theta_{y+1,z}\\
	&+(n_1\theta_{y,z-1}+n_2\theta_{y,z+1}+n_2\theta_{y+1,z-1}+n_1\theta_{y+1,z+1}).
\end{split}
	\label{eqn:coupling}
\end{equation}
This is a generalization of the coupled wire construction for bilayer quantum Hall states by Teo and Kane \cite{TeoKane2014}. The interaction $\cos\Theta$ is illustrated in Fig.~\ref{fig:couple-wire}(b). One can easily show that these fields satisfy the null vector condition\cite{Haldane1995}:
\begin{equation}
    [\Theta_\mb{r}(x), \Theta_{\mb{r}'}(x')]=0,
\end{equation}
so they can be minimized simultaneously. 
 We are interested in the strong-coupling limit $g\rightarrow \infty$, at which the cosine terms pin all $\Theta$ fields to the minima. This can be achieved either by having a large bare value of $g$, or turning on inter-wire density-density interactions to make $g$ a relevant coupling.

First we study the global symmetry of the model. Before turning the coupling Eq. \eqref{eqn:coupling}, each wire has $\U_\varphi\times\U_\theta$ symmetry, the charge densities of which are $\frac{1}{2\pi}\partial_x\theta$ and $\frac{1}{2\pi}\partial_x\varphi$, respectively, and there is a mixed 't Hooft anomaly between the two $\U$ subgroups. One of them, $\U_\varphi$, can be understood as the conservation of boson number, which is assumed to be an exact symmetry of the microscopic Hamiltonian. Here we choose the convention that the boson density is given by $\frac{1}{2\pi}\partial_x\theta$, and the symmetry transformation acts on the fields as
\begin{equation}
    \varphi\rightarrow \varphi+\alpha,
\end{equation}
where $\alpha\in [0,2\pi)$ is the $\U$ rotation angle.
The other subgroup $\U_\theta$ then must be emergent at low energy due to the mixed anomaly. Importantly, the lattice translation symmetry along the wire is realized at low energy as a particular element of $\U_\theta$:
\begin{equation}
    T_x: \theta\rightarrow \theta+2\pi\nu,
\end{equation}
where $\nu$ is the filling factor of the wire, which is equal to the filling factor of the entire 3D system. When $\nu$ is not an integer, the celebrated Lieb-Schultz-Mattis theorem~\cite{LSM} shows that the 1D wire cannot be fully gapped without breaking the translation symmetry. 

The coupling Eq. \eqref{eqn:coupling} apparently reduces the symmetry group. For each $xy$ plane, there is a $\U_\varphi$ subsystem symmetry:
\begin{equation}
    \varphi_{yz}\rightarrow \varphi_{yz}+\alpha_z,
    \label{eqn:phiU1}
\end{equation}
However, the $\U_\theta$ symmetry is broken in general.
 
We are mainly interested in the special case when $m=-(n_1+n_2)$. For this choice of parameters,  we find a $\U_\theta$ subsystem symmetry in each $xz$ plane:
\begin{equation}
    \theta_{yz}\rightarrow \theta_{yz}+\beta_y.
    \label{eqn:thetaU1}
\end{equation}
This obviously includes a global transformation $\theta_\mb{r}\rightarrow \theta_\mb{r}+\beta$.

If we set $n_1=n_2$, then there is a further U(1) dipole symmetry on each $xz$ plane:
\begin{equation}
    \theta_{yz}\rightarrow \theta_{yz}+z\gamma_y.
\end{equation}

We now consider spatial translations. The full Hamiltonian is obviously invariant under the discrete translations along $y$ and $z$ directions. For the lattice translation $T_x$, the interaction is invariant if and only if $(m+n_1+n_2)\nu$ is an integer. If $m=-(n_1+n_2)$, then $\nu$ can be any real number. Otherwise the model is only translation-invariant for certain special rational fillings.  We thus conclude that the model with $m=-(n_1+n_2)$ can be defined \emph{at any filling}. As will be shown below, there is no spontaneous symmetry breaking for either the global $\U_\varphi$ or $\U_\theta$, in the sense that no local order parameter exists. Therefore the coupled wire model constitutes a new example of compressible quantum phase.

To summarize, we find that at $m=-(n_1+n_2)$, the system preserves the following global continuous symmetry:
\begin{equation}
\begin{gathered}
    \U_\varphi: \varphi_\mb{r}\rightarrow \varphi_\mb{r}+\alpha,\\
    \U_\theta: \theta_\mb{r}\rightarrow \theta_\mb{r}+\beta.
\end{gathered}
\label{eqn:mixedU1}
\end{equation}
These two U(1) symmetries have a mixed 't Hooft anomaly, so cannot be realized as on-site symmetries microscopically at the same time. In this work we choose the convention that $\U_\varphi$ is the boson number conservation, and $\U_\theta$ is an emergent symmetry. The lattice translation along wires is embedded into $\U_\theta$.

\begin{figure}
\centering 
  \subfigure[]{\includegraphics[width = .49\columnwidth]{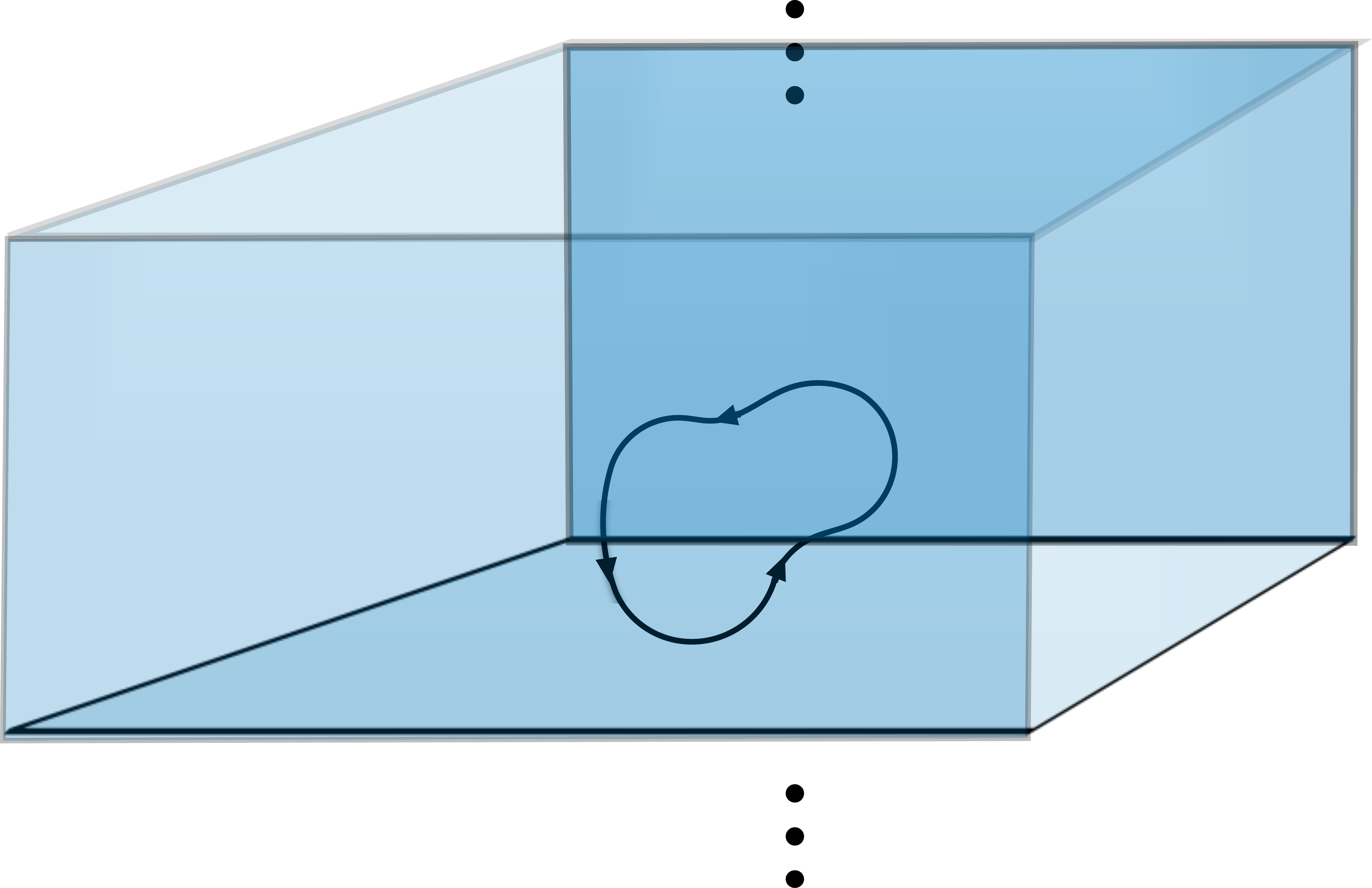}}
\subfigure[]{\includegraphics[width = .49 \columnwidth]{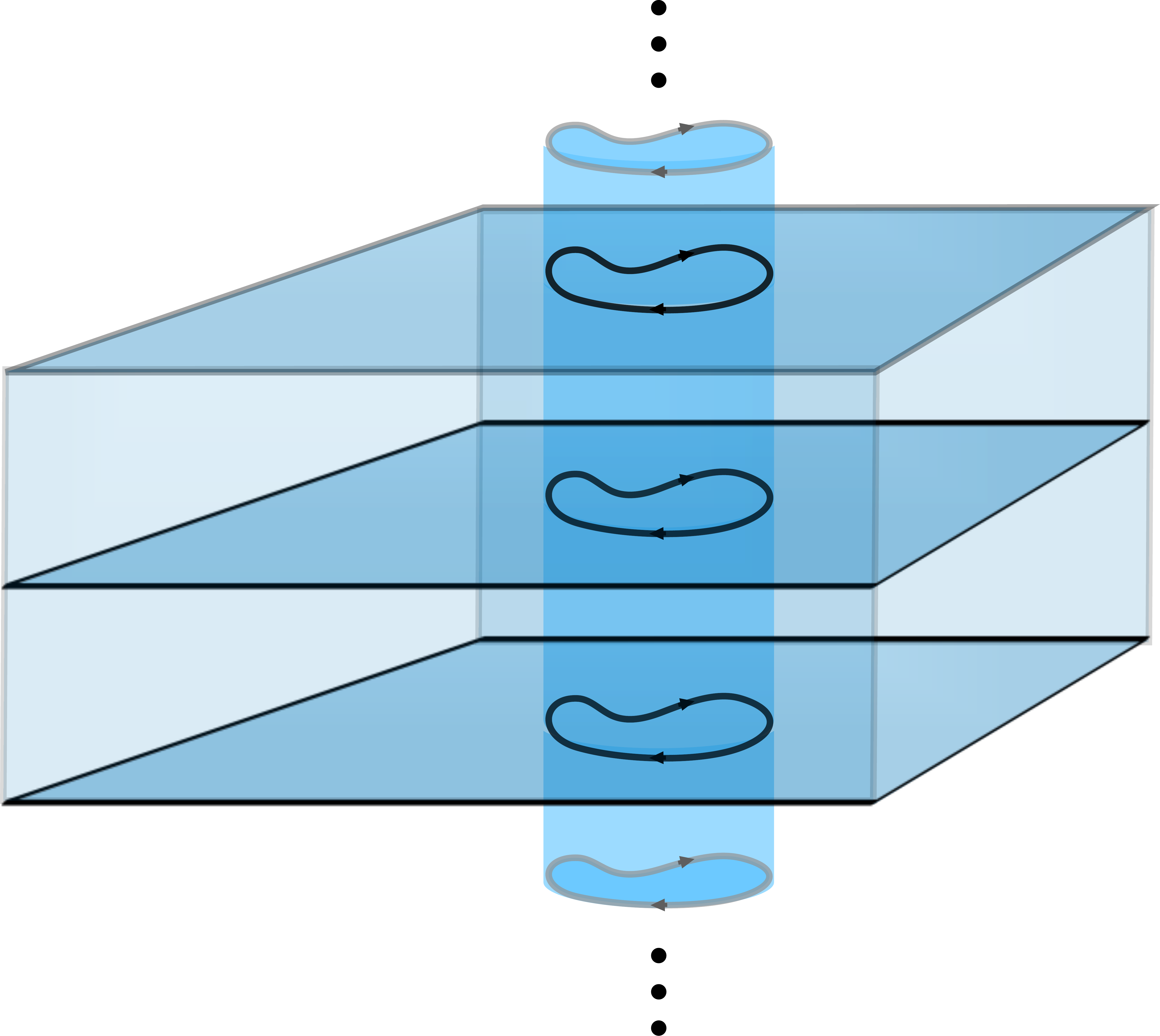}}
    \caption{(a) The 2-form symmetry operators in a 3D superfluid are supported on closed curves. The corresponding charged objects form closed surfaces (i.e. vortex sheets), hence a 2-form symmetry. (b) The higher-form symmetry operator of the WSB compressible liquid, which is defined in Eq \ref{eq:1-form charge}. The surface is deformable in $xy$-plane provided it is deformed in the same way in each layer. We refer to this as a cylindrical 1-form symmetry.}
    \label{fig:charge operators}
\end{figure}

So far we have only considered the so-called 0-form symmetries. At low energy, the system can develop emergent ``higher-form" symmetries~\cite{Gaiotto_2015}, whose charges are extended objects. An example relevant to our discussion is the U(1) 1-form symmetry in a 2D superfluid, which is physically equivalent to all superfluid vortices being non-dynamical (i.e. infinitely heavy). In our model, if $m=n_1=n_2=0$, the system is a stack of layers of 2D superfluids, and at low energy each layer has its own U(1) 1-form symmetry. Schematically, the 1-form charge in the layer $z$ for a closed path $C$ is given by
\begin{equation}
    Q_{z}(C)=\oint_C \nabla \varphi_z,
\end{equation}
which counts the total vorticity enclosed by $C$.
When the coupling to $\theta$ is turned on, since $e^{i\theta}$ creates a vortex-anti vortex pair and hops vortices in the layer, the U(1) 1-form symmetry for the layer is explicitly broken. However, when $m=-n_1-n_2$, the term $\cos (\Theta_{y,z})$ preserves the 1-form charge
\begin{equation}
    Q_{z-1}(C)+Q_z(C) + Q_{z+1}(C).
\end{equation}
Note that $C$ must be the same for the three layers. Now considering all layers together, the total 1-form charge
\begin{equation}
\label{eq:1-form charge}
    \sum_z Q_z(C)
\end{equation}
is conserved by the Hamiltonian at low energy when the cosine term dominates, assuming periodic boundary condition. We refer to the conservation of the total vorticity Eq. \eqref{eq:1-form charge} as a cylindrical 1-form symmetry, since the closed loop $C$ must be exactly aligned throughout all the layers, so the symmetry operator is in fact supported on a cylindrical surface. This should be compared with the 2-form symmetry in a 3D superfluid, where the symmetry operator is supported on arbitrary loops. We illustrate the symmetry operator in Fig \ref{fig:charge operators}b.

\subsection{Spectrum and excitations}
\label{sec:spec}
We now move on to analyze the spectrum of the Hamiltonian. The model is generally not solvable, so we limit ourselves to the strong-coupling limit where the cosine potential terms dominate over the kinetic terms. As shown in Ref. [\onlinecite{FractonCW2021}], low-energy excitations can be divided into two types: first of all, there are excitations that describe smooth fluctuations of the bosonic fields, similar to spin waves. They can be analyzed in the ``mean-field" approximation, where we expand $\cos \Theta_\mb{r}\approx 1-\frac12\Theta_\mb{r}^2$, and then solve the quadratic theory. We find that the spectrum of such Gaussian fluctuations is given by
 \begin{equation}
     \label{eq:gap equation}
     E_\mb{k} = \sqrt{v^2k_x^2 + vg(f_\varphi + f_\theta)}
 \end{equation}
 where 
 \begin{equation}
\begin{split}
f_\varphi &= 2-2\cos k_y,\\
f_\theta &=4\left[m \cos \frac{k_y}{2}+n_1\cos 
 (\frac{k_y}{2}-k_z)+n_2\cos (\frac{k_y}{2}+k_z)\right]^2.
\end{split}
\end{equation}
When $|m|\leq |n_1+n_2|$, one can make both $f_\varphi=f_\theta=0$ by setting $k_y=0, \cos k_z=-\frac{m}{n_1+n_2}$, which is clearly the minimum of $f_\varphi+f_\theta$ since both functions are non-negative. Therefore the spectrum is gapless.
 For $m=-(n_1+n_2)$, the minimum is at $\mb{k}=(0,0,0)$. Near this point, $E_\mb{k}$ takes the following approximate form:
\begin{equation}
\label{eq:Wire Spectrum}
\begin{split}
    E_\mb{k}&=\Big[v^2k_x^2+vg\Big(k_y^2 +2(n_2^2-n_1^2)k_yk_z^3\\
    &\quad\quad+(n_1-n_2)^2k_y^2k_z^2+(n_1+n_2)^2 k_z^4\Big)\Big]^{1/2}. 
    \end{split}
\end{equation}
For $n_1=n_2=n$, we have
\begin{equation}
    f_\varphi+f_\theta=4\cos^2\frac{k_y}{2}[(m+2n\cos k_z)^2-1]+4.
\end{equation}
If $|m|>2|n|$, the minimum of this function is 4 at $k_y=\pm\pi$ and the spectrum is fully gapped. When $|m|\leq 2|n|$, the minimum is $0$ with $k_y=0$ and $\cos k_z=-\frac{m}{2n}$. In this case the spectrum is gapless, with one or two gap-closing points. For $m=-2n$, near the gap-closing point $\mb{k}=(0,0,0)$, the spectrum is approximately
\begin{equation}
    E_\mb{k}=\sqrt{v^2k_x^2+vgk_y^2+4vgn^2 k_z^4}.
\end{equation}

Besides smooth fluctuations, there also exist localized excitations, corresponding to ``discontinuities" in the field configurations. At the mean-field level, they can be built out of soliton excitations of the gapping terms that tunnel between different minima. More concretely, for $\cos \Theta_\mb{r}(x)$, a $k$-soliton where $k\in\Z$ at $x_0$ is a configuration where $\Theta_\mb{r}$ winds by $2\pi k$ over a short distance $\xi$ around $x_0$. In this limit, these excitations can be thought of as massive quasi-particles.  While we do not know the exact profile of such an excitation, we assume that they stay gapped beyond the mean-field analysis, at least for a range of $g$. We give a more systematic description of the solitonic excitations in Sec. \ref{sec:qp}.

To create such excitations, one can apply vertex operators, such as $e^{i\varphi}$ or $e^{i\theta}$. Note that these two, together with derivatives of $\varphi$ and $\theta$, generate all local operators in the wire model. It is easy to show that derivatives of $\varphi$ or $\theta$ cannot create localized solitons. Rather they only result in smooth configurations of the fields. Therefore we can focus just on the vertex operators.  A crucial fact which we rely on here is that the model satisfies a version of the ``topological order" condition as discussed in Ref. [\onlinecite{FractonCW2021}].  Namely, if a vertex operator has a finite support and commutes with all $\Theta_{\mb{r}}$, then it must be a linear combination of $\Theta_{\mb{r}}$'s. In other words, any vertex operator with local support has to create gapped excitations, provided it is not some combination of $\Theta_{\mb{r}}$'s. Therefore such operators are infinitely irrelevant even when the theory is gapless. This strongly suggests that the gapless theories, when $|m|\leq|n_1+n_2|$, are stable with respect to weak local perturbations, as long as the topological order condition is obeyed. This is analogous to 3D U(1) gauge theory with gapped charges.

It is important to note that the topological order condition is shown for an infinitely large system. When the system is finite (or finite in one direction), the situation turns out to be quite different. As we discuss next, there can be vertex operators commuting with all gapping terms. Such operators are crucial in understanding the symmetry breaking (or the absence thereof) in the low-energy theory.

\subsection{String superfluid order}
\label{sec:stretched superfluid order}
In this section we focus on $m=-(n_1+n_2)$ and assume that both global $\U_\theta$ and $\U_\varphi$ symmetries are preserved. The topological order condition does not preclude the existence of non-local vertex operators that commute with all plaquette terms. We can consider a closed system with periodic boundary condition along $z$, in which case we can identify two such non-local operators: 
\begin{equation}
 \Phi_y = \sum_z \varphi_{yz}, ~ \Pi_z=\sum_{y}\theta_{yz}.\end{equation}
Note that when $n_1=n_2$ there is an additional operator of interest given by $D_y=\sum_z z\varphi_{yz}$, but it does not exist with PBC. The operator $e^{i\Phi_y}$ has a charge $N_z$ under $\U_\varphi$, where $N_z$ is the size of system in the $z$ direction. Likewise, $e^{i\Pi_z}$ has a charge $N_y$ under $\U_\theta$.

We show in Appendix \ref{sec:order_parameter} that with a finite $N_z$, when viewed as a quasi-two-dimensional system, the $e^{i\Phi_y}$ operator orders and spontaneously breaks the $\U_\varphi$ symmetry. More precisely, the two-point function decays as
\begin{equation}
    \langle e^{i\Phi_y(x)}e^{-i\Phi_{y'}(x')}\rangle\rightarrow e^{-\frac{L_z}{\rho |\mb{r}-\mb{r}'|}},
\end{equation}
where $\rho$ is a constant, and $\mb{r}=(x,y)$ is the coordinate in the xy plane.
Therefore, fixing $N_z$, the correlation function approaches a constant when $|\mb{r}-\mb{r}'|$ is much greater than $N_z/\rho$. This makes sense since for such large separation the thickness along the $z$ direction becomes negligible. However, going to the 3D limit when $N_z$ is comparable to the size of the other dimensions, the superfluid order is suppressed and eventually vanishes in the thermodynamic limit $N_z\rightarrow\infty$.

Equivalently, one can consider the superfluid stiffness in the quasi-two-dimensional system. The ``Josephson" coupling  that stabilizes the superfluid order originates from the $\varphi_{y,z}-\varphi_{y+1,z}$ term in $\Theta_\mb{r}$. However, such a term only accesses a small local portion of the order parameter, which is highly non-local and runs through the entire $z$ direction.  As a result, one can show that the superfluid stiffness vanishes as $1/N_z$.

{To summarize, we find that the line operator $e^{i\Phi_y}$ develops long-range order breaking the U(1)$_\varphi$ symmetry, but there exist no local U(1)$_\varphi$ order parameter. This is what we refer to as the phenomenon of ``weak symmetry breaking".
We note that similarly $e^{i\Pi_z}$ exhibits long-range order in the $x-z$ plane, although the asymptotic of the correlation function is now more anisotropic.} So the U(1)$_\theta$ symmetry is also weakly broken by the line order parameter.

It is also instructive to consider open boundary conditions along $z$. Suppose there are two $yz$ surfaces at $z=0$ and $z=N_z-1$. We first choose the gapping terms to be 
\begin{equation}
    -g\sum_y\sum_{z=1}^{N_y-2} \cos \Theta_{yz}.
\end{equation}
At each $y$, there are $N_z$ wires but only $N_z-2$ gapping terms, which leaves certain fields near the boundary unpinned. We focus on the $z=0$ surface. Unpinned fields on the surface are generated by $\theta_{y0}$ and 
\begin{equation}
\label{eq:surface dof}
    \tilde{\varphi}_{y+\frac12} = \varphi_{y+1,0}-\varphi_{y,0}+n_1\theta_{y,1}+n_2\theta_{y+1,1}.
\end{equation}
Note that both $\tilde{\varphi}_{y+\frac12}$ and $\theta_{y0}$ commute with themselves.
We see that all unpinned fields are neutral under $\U_\varphi$ symmetry.

Now we further add interactions on the surface, e.g. $\cos (\tilde{\varphi}_{y-\frac12}-\tilde{\varphi}_{y+\frac12})$ or $\cos(\theta_{y0}-\theta_{y+1,0})$, which spontaneously break the $\U_\theta$ symmetry. In this case, the $e^{i\Phi_y}$ operator no longer commutes with the surface order and thus is not a low-energy observable.

We can also create a completely symmetric and gapped surface, by adding the following perturbations:
\begin{equation}
\label{eq:symmetric surface term}
    -\cos \left[\frac{m}{2}(\theta_{y,0}+\theta_{y+1,0})+\tilde{\varphi}_{y+\frac12}\right].
\end{equation}

One can check that all the cosine arguments commute with each other and there are enough of them to gap out all the degrees of freedom on the surface. In this case, one can slightly modify the definition of $\Phi_y$ at the surfaces so that it still commutes with all the gapping interactions. This is done by adding a vertex operator which generates an appropriate phase slip at the end of the string operator: $\Phi_y \to \sum_z \varphi_{yz} - (\frac{n_1-n_2}{2})\theta_{y0}$. Note that when $n_1=n_2=-m/2$, $\Phi_y$ has the same form as the PBC case. 

While we do not have a general proof, we believe one of following scenarios must occur: either the $\U_\theta$ symmetry is spontaneously or explicitly broken at the surface, in which case no (local or non-local) $\U_\varphi$ order parameter exists in the system, or the $\U_\theta$ symmetry is preserved but one can then find a non-local order parameter by modifying $\Phi_y$ near the surface.

Since the model has a non-local superfluid order, one may wonder whether there is any superconducting response once coupled to an external electromagnetic field. Given the model is highly anisotropic, it is not surprising that the answer depends on direction. As shown in Eq. \eqref{eqn:phiU1}, the U(1) charge in each layer is conserved by the Hamiltonian as well as the ground state. As a result there cannot be any charge transport in the $z$ direction, i.e. it is insulating. The other directions are quite different. Using the dual gauge theory we compute the effective action for the external field in Appendix \ref{sec:Electromagnetic response}, and indeed show that the system is superconducting in the $xy$ plane.


\subsection{Duality mapping}
\label{sec:Duality mapping}
In this section we describe how the microscopic wire model can be mapped to an infinite-component CS theory coupled to gapped matter fields. The full details of this procedure are presented in Appendix \ref{sec:Details of Duality Mapping}.
 Schematically, the gauge theory emerges by employing a boson-vortex duality within each horizontal layer of wires. We adopt the method explained in  Refs. [\onlinecite{mrossPRX2017},\onlinecite{LeviatanPRR2020}] to the 3D coupled wire model. Note that an alternative method to derive Chern-Simons-type gauge theory from coupled wire models have been proposed in Ref. [\onlinecite{HanssonCW}].

Recall that the wire model is described by a rectangular array of Luttinger liquids with conjugate variables $(\varphi, \theta)$ and Lagrangian
\begin{equation} 
\label{eq:WireLagrangian1} 
\begin{split}
\mathcal{L}[\varphi,\theta]&=\sum_\vr \frac{i}{\pi}\partial_x \theta_\vr \partial_\tau \varphi_\vr+\frac{\tilde{v}}{2 \pi}\left(\partial_{x} \varphi_{\vr}\right)^{2}+\frac{u}{2 \pi}\left(\partial_{x} \theta_{\vr}\right)^{2}\\
&~~~~~~+\frac{v}{8 \pi}\left(\partial_{x} \Delta_y \varphi_{\vr}\right)^{2} - g\cos(2\bth_\vr)
\end{split}
\end{equation} 
where 
\begin{equation}
\label{eq:Plaquette def}
    \begin{split}
        2\bth_{\vr} &= \Delta_y \varphi_\vr + \Lambda \theta_\vr\\
        &=\left(\varphi_{\vr+\hat{y}}-\varphi_{\vr}\right) \\
        &~~~~+(m \theta_{\vr} +m\theta_{\vr+\hat{y}} +n_1\theta_{\vr-\hat{z}}+n_2\theta_{\vr+\hat{z}}\\
        &~~~~+n_2\theta_{\vr+\hat{y}-\hat{z}}+n_1\theta_{\vr+\hat{y}+\hat{z}}).
\end{split}
\end{equation}
Here we define $\Delta_y X_\vr = X_{\vr+\hat{y}}-X_r$ and $\vr=(y,z)$ is a wire index.  Note that the kinetic part of Eq. \eqref{eq:WireLagrangian1} is more general than the free boson Hamiltonian given in Eq. \eqref{eq:free boson H}. In particular, $\frac{v}{8 \pi}\left(\partial_{x} \Delta_y \varphi_{\vr}\right)^{2}$ has been added to the standard Luttinger Liquid kinetic term. This is done because it is convenient for the duality mapping to the vortex theory but its presence does not affect the qualitative physics \cite{mrossPRX2017}.
 
Throughout the discussion of the microscopic model we use two equivalent forms of labelling for the wires, $O_{\vr+\hat{a}} \equiv O_{y+a_y,z+a_z}$. Anticipating the layered structure of the gauge theory, we treat $z$ as the ``layer" index while $y$ is coarse grained to a continuous spatial coordinate. With this motivation in mind, we define the following pair of conjugate variables: 
\begin{equation}
\label{eq:vortexvar}
\begin{array}{l}
\tph_{yz}=-\sum_{y^{\prime}} \sgn\left(y^{\prime}-y-\frac{1}{2}\right) \theta_{y^{\prime}z} \\
\\
\tth_{yz}=\frac12\left(\varphi_{y+1,z}-\varphi_{y,z}\right) . 
\end{array}
\end{equation}
Here $\tph_{yz}$ creates a $2\pi$ vortex in the $\varphi$ field in layer $z$, in between wires $y$ and $y+1$. The operator $\partial_x \tth_{yz}$ is the ``charge" operator for this vortex.
We can re-express the Lagrangian in Eq. \eqref{eq:WireLagrangian1} in terms of these new fields but the result will be  highly non-local in the $y$ direction. We can restore locality in each layer $z$ via a Hubbard-Stratonovich transformation at the expense of adding new degrees of freedom $a_0^{(z)}(x,y)$ and $a_1^{(z)}(x,y)$.
At this stage, the Lagrangian takes the form
\begin{equation}
\label{eq:Vortex_Lag}
\begin{split}
\mathcal{L}[\tph,\tth,a_\mu] &= \sum_r \frac{i}{\pi} \partial_{x} \tilde{\theta}_\vr (\partial_\tau \tph_\vr -a_{0,\vr}^{(z)}) + \frac{u}{2 \pi}\left(\partial_{x} \tph_{\vr}-a_{1,\vr}^{(z)}\right)^{2}\\
&~~~~~~+\frac{v}{2 \pi}\left(\partial_{x} \tth_\vr\right)^{2} - g\cos(2\tth_\vr +\frac{1}{2}\Lambda \cdot \Delta_y \tph_\vr)\\
&~~~~~~+ \mathcal{L}_{\text{Maxwell}}^{(z)}.
\end{split}
\end{equation}
Here, the definition of $\Lambda$ can be easily inferred from Eq. \eqref{eq:Plaquette def}, so that the argument of the cosine term matches $2\bar{\theta}_\mb{r}$. $\mathcal{L}_{\text{Maxwell}}^{(z)} = \frac{u}{2\pi}(\Delta_y a_1^{(z)})^2 + \frac{1}{2\pi \tilde{v}}(\Delta_y a_0^{(z)})^2$. The superscript of this term is no accident; we interpret Eq. \eqref{eq:Vortex_Lag} as a stack of 2D gauge theories in the $a_2^{(z)} = 0$ gauge coupled to matter $(\tph,\tth)$. In the context of Eq. \ref{eq:Vortex_Lag} the gauge fields would appear to be over-labeled since $\vr = (y,z)$. We keep the redundant $(z)$ superscript to emphasize this picture of stacks of 2D gauge theories. Such ``layered" gauge theories have recently been studied in the context of fracton physics \cite{ma2020fractonic} and are reminiscent of the foliated gauge theories developed to describe fracton order \cite{SlagleFoliated}. We emphasize that the interpretation of $a$'s as a stack of 2D gauge fields in the $xy$ plane is motivated, and ultimately justified by providing a correct description of the low-energy physics of the model.

The next step is to integrate out the fields $\tilde{\varphi}$ and $\tilde{\theta}$. The details of the derivation can be found in  Appendix \ref{sec:Details of Duality Mapping}. In the end we obtain the effective layered Maxwell-CS theory:
\begin{equation}
    \mathcal{L}[a] \equiv \frac{iK_{zz^\prime}}{4\pi} a^{(z)} \wedge da^{(z^\prime)} + (\text{Maxwell terms}),
    \label{eqn:maxwell-cs}
\end{equation}
where 
\begin{equation}
  K_{zz^\prime} = 2m\delta_{zz^\prime} + (n_1+n_2)\delta_{z, z^\prime\pm 1}.
  \label{eqn:Kmat}
\end{equation}
The inter-layer Maxwell term contains couplings between components of the electromagnetic field from adjacent layers $(z)$; as an example the $B^2$ term corresponds to $\sum_{z,z^\prime} B^{(z)}\left[ \frac{u}{2\pi}\delta_{zz^\prime} + \frac{v}{8\pi}(K^\mathsf{T}K)_{zz^\prime}\right]B^{(z^\prime)}$.

So far we have assumed that $\varphi$ and $\theta$ vary smoothly, excluding the ``solitonic" excitations of the cosine pinning term. These more singular configurations are the analog of vortices in the usual boson-vortex duality, which should be minimally coupled to the dynamical gauge fields. In Appendix \ref{sec:Coupling to vortices} we show that this is indeed the case; namely the solitons become gauge charges in the dual gauge theory. 


\subsubsection{OBC in z direction}
\label{sec:obc}
Now we consider OBC in the $z$ direction and the effect of the various surface gapping terms discussed in Sec \ref{sec:stretched superfluid order} on the resulting dual gauge theory.

First we discuss the symmetrically gapped surface generated by the pinning term in Eq. \eqref{eq:symmetric surface term}. The analysis is similar to the PBC case, except now we must pay attention to the $z=0$ layer of wires. We decompose $\sum_\mb{r}$ into $\sum_y\sum_{z>0} + \sum_{y,z=0}$ in Eq \eqref{eq:WireLagrangian1}. The terms in the first sum are unchanged while in the second sum, $\sum_{y,z=0}$, the surface pinning term is $\cos 2\bth_{y0}$ where
\begin{equation}
\begin{split}
2\bth_{y0} &= 2\tth_{y0} + \frac{m}{4}\Delta_y (\bphi_{y,0}+\bphi_{y+1,0})\\
&~~~~~~~~~~~~~~~~+ \frac{n_1}{2}\Delta_y \bphi_{y,1}  + \frac{n_2}{2}\Delta_y \bphi_{y+1,1} ~~~.
\end{split}
\end{equation}
From here one can integrate out the matter fields and truncate higher order derivative terms where appropriate. The upshot of this is a modification of the K-matrix:
\begin{equation}
    \label{eq:K OBC symmetric}
    \mb{K}_{\text{sym,OBC}}=m\begin{pmatrix}
    -1 & 1 & & & 0 \\
    1 & -2 & 1 & & \\
    & \ddots & \ddots & \ddots & \\
    & & 1 & -2 & 1 \\
    0 & & & 1 & -1 
    \end{pmatrix}.
\end{equation} 
The $B^2$ term is also modified near the surface. This can be seen by noting that the term in the Lagrangian corresponds to something proportional to $(1+\alpha K^\mathsf{T}K)_{zz^\p}B^{(z)}B^{(z^\p)}$ for some constant $\alpha$.

Alternatively, we can explicitly break the U(1)$_\theta$ symmetry by adding the very natural pinning term corresponding to the middle and top layer of the bulk plaquette term. This partial plaquette is given by 
\begin{equation}
\begin{split}
2\bth_{y0} &= 2\tth_{y0} + \frac{m}{2}\Delta_y (\bphi_{y,0}+\bphi_{y+1,0})\\
&~~~+ \frac{n_1}{2}\Delta_y \bphi_{y,1}  + \frac{n_2}{2}\Delta_y \bphi_{y+1,1}.
\end{split}
\end{equation}
The duality mapping of this surface theory proceeds in the same way as the symmetric gapped example just discussed. In this case 
\begin{equation}
    \label{eq:K OBC symmetric}
    \mb{K}_{\text{broken,OBC}}=m\begin{pmatrix}
    -2 & 1 & & & 0 \\
    1 & -2 & 1 & & \\
    & \ddots & \ddots & \ddots & \\
    & & 1 & -2 & 1 \\
    0 & & & 1 & -2 
    \end{pmatrix}.
\end{equation} 

\subsection{Vortices and quasiparticles}
\label{sec:qp}
Throughout our derivation of the dual gauge theory, we have assumed that the bosonic fields $\varphi$ and $\theta$ vary smoothly in spacetime. As we have already mentioned, the complete spectrum also contains solitonic configurations. 
We show in Appendix \ref{sec:Coupling to vortices} that the solitons are in fact dynamical charges of the dual gauge fields.  Below we describe the universal properties of the soliton excitations. Many details are delegated to Appendix \ref{sec:Mobility of solitonic excitations}.

Physically, a ``fundamental" soliton, i.e. a $2\pi$ jump in a single $\Theta_\mb{r}$ term, is in fact a vortex of the weak superfluid. This can be demonstrated by explicitly constructing a string  operator to create a pair of such solitons separated in the $y$ direction. The string operator takes the following form:
\begin{equation}
    \prod_y e^{i\theta_{yz}(x)},
    \label{eqn:y-string}
\end{equation}
which causes a branch cut of $2\pi$ phase jump in the string order parameter, thus the corresponding excitation is a vortex.
It is useful to divide the excitations into vortices and non-vortices. The simplest non-vortex excitation is just a dipole of two vortices, with opposite vorticities $k$ and $-k$, separated along the $z$ direction. Such non-vortex excitations can be considered as being ``deconfined" quasiparticles in this phase. We call $k$ the strength of the dipole.

Now we discuss the mobility of excitations. The string operator Eq. \eqref{eqn:y-string} shows that all excitations can move freely along the $y$ direction. In Appendix \ref{sec:Mobility of solitonic excitations} we show that such a fundamental soliton is immobile in the $z$ direction, when $|m|\geq 2|n|$. Notice that these results hold true in both gapped and gapless models. For the gapless theory with $m=-2n$, one can show that a dipole whose strength is a multiple of $m$ can move in the $z$ direction.

The question of mobility along the $x$ direction is more subtle. We present a construction of a $x$-string operator in Appendix. The string operator is ultra-local in the $y$ direction, but can have some extension in the $z$ direction. For a gapped model, we find that the construction does yield an exponentially localized string operator, as expected. However, the gapless case is quite different. To be concrete let us consider a finite $N_z$ system. For a vortex, the construction fails if PBC or the symmetric OBC is imposed in the $z$ direction. Only when the symmetry-breaking OBC is present one can find a string operator. For a dipole, a string operator always exists regardless of the boundary condition in the $z$ direction. In both cases, however, the constructed string operator decays only algebraically in the $z$ direction away from the localization of the excitation. This observation suggests that the mobility of these excitations along the $x$ direction is also reduced.  

In a gapped system, one can define the superselection sectors of quasiparticle excitations as the equivalence classes under local operations. Namely, two excitations are equivalent if they can be transformed to each other by local operator.  We can attempt to generalize the notion to our model. We focus on the non-vortex excitations, because there are infinitely many types labeled by the vorticity. Since every non-vortex excitation is mobile along $y$, we can just consider them at a fixed $y$. Equivalence between excitations located at different $x$ is less clear since the string operator is only algebraically localized. If however we consider them to be equivalent,  we find that there are $m^{N_z-1}\cdot N_z$ classes of non-vortex excitations when PBC is imposed, which agrees with the topological ground state degeneracy of the coupled wire model. More details of the counting can be found in Appendix \ref{sec:Mobility of solitonic excitations}.

Since the vortices are coupled to the gapless gauge fields, gauge fluctuations induce long-range interactions between the vortices. We compute the interactions when the simplest, most isotropic Maxwell term is added to the dual CS theory (see Eq. \eqref{eqn:maxwell2} below). For two unit vortices separated by $\mb{r}=(x,y,z)$, let $\rho=\sqrt{x^2+y^2}$. The interaction potential takes the following asymptotic form:
\begin{equation}
    V(\mb{r})\sim\begin{cases}
    \frac{1}{|z|} & \tilde{g}z^2\gg \rho\\
    \frac{1}{\sqrt{\rho}} & \tilde{g}z^2\ll \rho
    \end{cases}.
\end{equation}
From this one can further obtain the interactions between dipoles, which decay with a higher power. Details of the calculations can be found in Appendix \ref{sec:vortex interaction}.

\section{Infinite-component CS theory}
\label{sec:Layered CS gauge theory}
Our duality mapping suggests that the low-energy physics can be described by an infinite-component Chern-Simons gauge theory, {as given in Eq. \eqref{eqn:maxwell-cs}. The gapped case $|m|>|n_1+n_2|$ was studied in greater detail in Ref. [\onlinecite{ma2020fractonic}]. Briefly, such a theory exhibits a kind of fracton topological order, with all quasiparticles being planons. Interestingly, in many cases the Wilson loop operators for these quasiparticles must have exponentially decaying tails in the $z$ direction, and as a result, the braiding statistics between quasiparticles are not strictly local. When this phenomenon happens, the fractonic order appears to be beyond the topological defect network framework~\cite{AasenPRR2020, wang2020nonliquid, WenPRR2020}. }

{In the following we focus more on the gapless case, and our main task is to analyze the symmetry, anomaly and the issue of stability. We show that the iCS theory indeed captures the universal aspects of the coupled wire model. First of all, we note that the photon spectrum of the iCS theory with only ``intra-layer" Maxwell terms was calculated in Ref. [\onlinecite{ma2020fractonic}] and the result agrees \footnote{At first glance the spectrum of the layered gauge theory arising from Eq. \eqref{eq:Layered Spectrum} would seem to depend only on $n_1+n_2$ and $m$, which does not agree with the spectrum derived for the wire model. This discrepancy stems from the fact the $y$ direction has been coarse grained in the layered picture. Restoring higher order derivative terms in $y$ in the gauge theory gives a spectrum matching Eq. \eqref{eq:Wire Spectrum} } with the Gaussian spectrum Eq. \eqref{eq:Wire Spectrum} of the coupled wire model. Below we discuss how other universal aspects of the coupled wire model are encoded in the iCS field theory.
}

While we are interested in the iCS theory as a (3+1)d system, we start by reviewing the global symmetry and anomaly of a general $\U^{N_z}$ CS theory, which can be thought of as compactifying the 3D system in the $z$ direction (so there are $N_z$ ``layers"). {i.e. it reduces to a quasi-2D theory for small compactification radius or by allowing non-locality along the compactified direction}.

Notice that below we use differential form  notations extensively.

\subsection{Global symmetry and anomaly}
\label{sec:symmetry and anomaly}
We consider the action of a $\U^N$ CS theory:
\begin{equation}
    \mathcal{S}=\int_{M_3}\frac{K_{IJ}}{4\pi}a_I\wedge da_J,
    \label{eqn:cs}
\end{equation}
where $M_3$ is a closed three manifold. $\mb{K}$ is a symmetric integer matrix. First we consider the theory with the Chern-Simons term only and assume that there are no matter fields. This is a valid assumption since all vortices are gapped, so well below the gap we can just study the pure gauge theory. 

\subsubsection{Global symmetry}
We now enumerate all the unitary symmetries, including 0-form and 1-form, of the topological action.

First consider the pure topological action Eq. \eqref{eqn:cs}. The theory has the following discrete symmetries:
\begin{equation}
    a_I\rightarrow W_{IJ}a_J,
\end{equation}
where $\mb{W}$ belongs to GL$(N, \Z)$, i.e. an invertible $N\times N$ integral matrix, such that $\mb{W}^\TT \mb{K} \mb{W}=\mb{K}$. Physically, when $\mb{K}$ is non-degenerate, $\mb{W}$ corresponds to a permutation of anyon types in the (2+1)d Abelian topological phase. 

In addition, a U(1)$^N$ gauge theory also has U(1)$^N$ 0-form symmetries, the conservation of magnetic fluxes. The currents of the symmetries are $\star da_I$, where $\star$ is the Hodge star operator. When $\mb{K}$ is non-degenerate, these symmetries are embedded into the 1-form symmetry group.

We now discuss the 1-form symmetry of the model, first assuming that $\mb{K}$ is invertible. A general 1-form transformation takes the following form:
\begin{equation}
    a_I\rightarrow a_I + q_I \lambda,
\end{equation}
where $\lambda$ is a properly normalized flat connection, i.e. $d\lambda=0$, and the holonomy of $\lambda$ along any 1-cycle is an integer multiple of $2\pi$. $q_I$ is an arbitrary real number at this point. When $q_I\in \Z$, the transformation can be viewed as a (possibly singular) gauge transformation. Additionally, one can prove that in order for the partition function of the theory to be invariant, the $q_I$'s have to satisfy the following quantization condition:
\begin{equation}
    K_{IJ}q_I\in\Z.
\end{equation}
The derivation can be found in Appendix \ref{sec:quantization}.
If $K$ is invertible, it implies that $q_I$'s must take the values
\begin{equation}
    q_I=(K^{-1})_{IJ}k_J, k_J\in \Z,
\end{equation}
and the 1-form symmetry forms a finite Abelian group, the determinant group of $\mb{K}$. Each 1-form transformation corresponds to an integer vector $\mb{m}$, modulo those that can be written as $\mb{K}\mb{l}$ for $\mb{l}\in\Z^N$, and is in fact one-to-one correspondent to anyon types. This is well-known, as the 1-form symmetries are generated by Wilson loops. The 1-form symmetry is generally anomalous if the corresponding anyon is not a boson~\cite{Hsin_2019}.

When $\mb{K}$ is degenerate, the null space of $\mb{K}$ needs to be considered separately. Suppose that $\mb{v}\in\Z^N$ is a null vector, i.e. $\mb{K}\mb{v}=0$. Without loss of generality we can assume $\mb{v}$ is primitive (i.e. $\mathrm{gcd}(v_1,v_2,\dots)=1$). In that case, the following 1-form transformation
\begin{equation}
    \mb{a}\rightarrow \mb{a} + \alpha\mb{v}\lambda
    \label{eqn:U1-1-form-CS}
\end{equation}
is always an exact symmetry of the 4d action (without any $2\pi\Z$ shift), where $\alpha$ is an arbitrary real number in $[0,1)$. Therefore each null vector of $K$ gives rise to a U(1) 1-form symmetry group. Physically, a null vector corresponds to a mode of the gauge fields without CS term, which can be dualized to a superfluid, which has a U(1) 1-form symmetry in the absence of vortices~\cite{Delacretaz_2020}.  More generally, if the dimension of the null space is $r$, then we have $\U^r$ 1-form symmetry.

Now we consider what happens when Maxwell terms are included. The $\U^N$ and 1-form symmetries are not affected at all by the Maxwell terms. The Maxwell terms may break the 0-form symmetry, however. We then have only the subgroup that preserves the Maxwell terms as well.

We make connections between the symmetries identified in the CS gauge theory, especially with the K matrix given by Eq. \eqref{eqn:Kmat}, and those found in the coupled wire lattice model which are discussed in Sec \ref{sec:Model Hamiltonian and symmetries}. The magnetic $\U$ 0-form symmetry of the $I$-th layer is nothing but the planar $\U_\varphi$ symmetry. The U(1) ``1-form" symmetry defined in Eq. \eqref{eqn:U1-1-form-CS} can be identified with the conservation of total vorticity Eq. \eqref{eq:1-form charge}. Finally, as we will see shortly, we can identify the string order parameter with a ``string" monopole operator.

\subsubsection{'t Hooft anomaly}
The emergent symmetry group defined in the previous section generally has 't Hooft anomalies. The discrete part of the 0-form symmetry group has to be analyzed on a case-by-case basis. The anomaly of the finite 1-form symmetry group was studied in Ref. [\onlinecite{Hsin_2019}], and since it is not particularly relevant to us we do not go into details. So in the following we study the mixed anomaly between the 0-form $\U^N$ and the 1-form symmetry group $\mathcal{A}\times \U^r$.

We consider, without any loss of generality, a U(1) 0-form symmetry with charge vector $\mb{t}$. In other words, the U(1) current is given by
\begin{equation}
    j=\sum_I t_I \star\! da_I.
\end{equation}
 We turn on a general background gauge field $A$ for the U(1) symmetry:
\begin{equation}
    \frac{t_I}{2\pi}\int a_I \wedge dA.
\end{equation}
To check whether this coupling can be compatible with the 1-form symmetries, perform a general 1-form gauge transformation $a_I\rightarrow a_I+q_I \lambda_I$ where now $\lambda_I$ is allowed to be non-flat. The action changes by $-\frac{q_It_I}{2\pi}\int A\wedge d\lambda$. To restore gauge invariance, the additional term can be cancelled by inflow from a 4d action
\begin{equation}
  S_\text{bulk}=\frac{q_It_I}{2\pi}\int_{M_4} B\wedge dA,  
\end{equation}
where $B$ is the background 2-form gauge field.

A similar argument works for the $\U^r$ part of the 1-form symmetry group: under the 1-form gauge transformation $a_I\rightarrow a_I+\alpha{v}_I\lambda$, to maintain gauge invariance we need to have a bulk theory given by
\begin{equation}
\label{eq:S bulk}
    S_{\text{bulk}}=\frac{\mb{v}\cdot \mb{t}}{2\pi}\int_{M_4}B\wedge dA,
\end{equation}
where now $B$ is a 2-form U(1) gauge field, transforming under the 1-form gauge transformation as $B\rightarrow B+\alpha d\lambda$.

 \subsection{The filling anomaly}

We now analyze the filling anomaly in the layered CS gauge theory. We are mainly interested in K matrices given in Eq. \eqref{eqn:Kmat}, although we present the analysis in a form that applies to more general cases. Such theories have a discrete translation symmetry, generated by
\begin{equation}
    T_z:a_I \rightarrow a_{I+1},
\end{equation}
provided that the K matrix satisfies
     $K_{IJ}=K_{I+1,J+1}$.
 This is the formal way of identifying the index $I$ as labeling layers in the $z$ direction. We assume that $\mb{K}$ is ``short-ranged" in the $z$ direction, which means that there exists an integer $d\geq 0$ such that $K_{IJ}=0$ if $|I-J|>d$. For now we assume that either $N_z\rightarrow \infty$, or periodic boundary condition is imposed so $a_{I+N_z}=a_I$.   This way, the layered CS theory describes a highly anisotropic 3D system.

We are going to show that the (3+1)d theory can exist at any filling. Instead of directly computing the 3D filling anomaly, we take a detour and study the compactified system, with periodic boundary condition imposed along the $z$ direction. Suppose that the 3D system has a filling factor $\nu$ (i.e. the average U(1) charge is $\nu$ per unit cell). When viewed as a quasi-2D system, the filling factor becomes $N_z\nu$. Here the dependence on $N_z$ reflects the 3D nature.

Before going to the details, we briefly review the theory of the filling anomaly. It has been understood now that even though the filling condition does not correspond to a true, quantized 't Hooft anomaly (essentially because the filling factor is a continuous quantity), its impact on the low-energy physics can be described in the same theoretical framework. We first present a formal argument, following the approach in [\onlinecite{SongPRR2021}].  The standard method to detect the 't Hooft anomaly of a global symmetry is to couple the system to the background gauge field of the symmetry, and compute the topological response. While translation is a spatial symmetry, at low energy it can be described effectively as an internal symmetry.  To this end, we formally introduce $\Z$ gauge fields $x^1$ and $x^2$ for the 2D translation symmetry. We also turn on a U(1) background gauge field $A$. The filling condition can be captured by the following response
\begin{equation}
 S_\text{bulk}=\nu\int dA\cup x^1 \cup x^2.
\end{equation}

Before getting to the CS gauge theory, we show how the filling anomaly can be embedded into the $\U^{[0]}\times \U^{[1]}$ 't Hooft anomaly discussed in Sec. \ref{sec:symmetry and anomaly}. 
Formally, we start from the bulk action
\begin{equation}
    \frac{k}{2\pi}\int B\cup dA.
\end{equation}
To realize the filling anomaly, we demand that the translation background gauge field is activated through the 2-form background: 
\begin{equation}
  B=2\pi\nu x^1\cup x^2.  
  \label{eqn:2formbg}
\end{equation}
Then we see that the action gives a filling factor $k\nu$.

We give a physical interpretation as follows. Eq. \eqref{eqn:2formbg} implies that when a loop charged under the U(1)$^{[1]}$ symmetry moves, it picks up a phase factor given by the flux of the background 2-form gauge field $B$ through the transverse area swept by the loop. Now recall that $x^{1,2}$ are translational gauge fields, $x^1\cup x^2$ can be heuristically interpreted as the area in the 2D plane. So Eq. \eqref{eqn:2formbg} basically says that a phase $2\pi \nu \cal{A}$, where $\cal{A}$ is the enclosed area, is attached to the loop. The anomaly action then implies that the loop is identified with magnetic field lines of the $\U^{[0]}$ gauge field. Therefore, the anomaly action and the particular background Eq. \eqref{eqn:2formbg} together can be summarized by the following intuitive picture: a $2\pi$ U(1)$^{[0]}$ flux picks up $e^{2\pi i \nu}$ phase factor when it moves around a unit area. This is exactly the Aharonov-Bohm phase expected from the filling factor.

Such a mechanism is realized by a (2+1)d superfluid.  A $2\pi$ $\U^{[0]}$ magnetic flux is trapped in a superfluid vortex. It is well-known that under boson-vortex duality, vortices see a background magnetic field, the strength of which is fixed by the filling factor. 

We can apply these considerations to the CS theory. The anomaly is given by Eq. \eqref{eq:S bulk}. 
 The charge vector is simply $t_I=1$, as we have shown in the duality mapping. We posit that the translation symmetry is realized via the $\U$ 1-form symmetry corresponding to the ``zero mode" $\mb{v}=(1,1,\cdots,1)^\mathsf{T}$, by the 2-form background given in Eq. \eqref{eqn:2formbg}. Together, we find that the coefficient of the topological action is $\mb{v}\cdot\mb{t}\nu=N_z\nu$. which is the expected filling factor. In fact, it is clear that as long as the charge vector $\mb{t}$ and $\mb{v}$ are translation-invariant (i.e. invariant under, or a certain multiple), the theory can be defined at arbitrary filling.
Physically, whenever a zero mode exists, after compactification the system becomes a quasi-2D superfluid where the U(1)$^{[0]}$ symmetry is spontaneously broken. 
 
 Notice that so far we have assumed periodic boundary condition in the $z$ direction. Now we turn to open boundary condition. In this case, the $z$ translation symmetry is broken. The filling of the quasi-2D system is still $N_z\nu$.  The boundary condition at the top and bottom layers now becomes crucial. We have considered two kinds of boundary conditions in the coupled wire model in Sec. \ref{sec:obc}, which lead to different K matrices after duality transformation.
If the boundary condition preserves all the symmetries, we found the following K matrix: 
\begin{equation}
    \label{eq:K_gl}
    \mb{K}_{\text{sym,OBC}}\propto\begin{pmatrix}
    -1 & 1 & & & 0 \\
    1 & -2 & 1 & & \\
    & \ddots & \ddots & \ddots & \\
    & & 1 & -2 & 1 \\
    0 & & & 1 & -1 
    \end{pmatrix}.
\end{equation}
With this K matrix, $(1,1,\dots,1)^\mathsf{T}$ is still a zero mode, and the same argument, as used for the periodic boundary condition, applies. 

We also studied a symmetry-breaking surface in Sec. \ref{sec:obc},  which yields the following K matrix:
\begin{equation}
    \label{eq:K_gl_sb}
    \mb{K}_{\text{broken,OBC}}\propto\begin{pmatrix}
    -2 & 1 & & & 0 \\
    1 & -2 & 1 & & \\
    & \ddots & \ddots & \ddots & \\
    & & 1 & -2 & 1 \\
    0 & & & 1 & -2 
    \end{pmatrix}.
\end{equation}
This K matrix is non-degenerate, thus representing a fully gapped 2D phase. As discussed in [\onlinecite{Gaiotto_2015}] and [\onlinecite{Hsin_2019}], such a CS theory has a finite emergent symmetry group, incompatible with a generic filling. Thus the only way that the theory can emerge is to to break the symmetries explicitly. This is indeed the case in the coupled wire model.

\subsection{Monopole operators}
\label{sec:monopole}
U(1) gauge theories generally admit monopole (or ``disorder") operators, which are charged under the magnetic U(1) symmetries. For $\U^N$ gauge group, a monopole operator is labeled by the magnetic charge vector $\mb{m}\in\Z^N$, i.e. $2\pi m_I$ flux for the $a_I$ gauge field. Physically, it is actually an instanton that inserts $2\pi\mb{m}$ flux at a point. Due to the Chern-Simons coupling, the flux insertion necessarily nucleates charges, so we expect that the monopole operator is not gauge-invariant.  In order to build a gauge-invariant monopole operator, one has to dress the ``bare" operator by matter fields of charge $\mb{K}\mb{m}$. As matter fields are absent (or very massive) in the low-energy theory, a gauge-invariant monopole operator can be introduced at low energy (i.e. well below the mass gap) only when $\mb{m}$ is a null vector of $\mb{K}$. For a translation-invariant K matrix, null vectors are also translation-invariant. 
They are precisely the string order parameters in the coupled wire model. Via the well-known Polyakov mechanism, adding  monopole operators to the theory causes confinement of the gauge theory. However, if the monopole operator has a translation-invariant magnetic charge vector, it is in fact a highly non-local, string-like object in 3D and cannot appear as a local term in the Hamiltonian. We thus conclude that such theories are stable with respect to confinement caused by monopole proliferation.

We now look closer at the monopole operators, which requires adding  Maxwell terms in the action:
\begin{equation}
    S_\mathrm{Maxwell}=\int \tilde{g}_{IJ}f_I\wedge \star f_J.
\end{equation}

For this purpose, we diagonalize the K matrix: 
\begin{equation}
  \sum_JK_{IJ}e_J^\alpha = \lambda^q e_I^q,  
\end{equation}
where $q$ labels different eigenvectors.
Here we normalize $e^q$ such that $\sum_Ie^q_I e^{q^\prime}_I=\delta^{q q^\prime}$. Then the K matrix can be written as $K_{IJ}=\sum_q \lambda^q e^q_I e^q_J$. Define new gauge fields $b^q = e_I^q a_I$ and the inverse transformation reads $a_I = e^q_I b^q$. In terms of the new fields, the CS term becomes
\begin{equation}
    \sum_q\frac{\lambda^q}{4\pi} b^q \wedge db^q.
\end{equation}
 and the Maxwell term becomes
\begin{equation}
    \tilde{g}_{IJ} e^q_I e^{q^\prime}_J f^q \wedge \star f^{q^\prime},
\end{equation}
here $f=db$.

For the simplest choice, we can set $ \tilde{g}_{IJ}= \tilde{g}\delta_{IJ}$, then $ \tilde{g}_{IJ}e^q_I e^{q^\prime}_J= \tilde{g}\delta^{q q^\prime}$. So the theory becomes
\begin{equation}
     \mathcal{L}=\frac{\lambda^q}{4\pi} b^q \wedge db^q -  \tilde{g} f^q \wedge \star f^q.
     \label{eqn:maxwell2}
\end{equation}
In our layered CS theory, the translation invariance along $z$ guarantees that different modes do not couple. In general, $ \tilde{g}^a$ stays finite even when $N\rightarrow \infty$, unless $ \tilde{g}_{IJ}$ is exactly proportional to $K_{IJ}$.

As we have already mentioned, the gauge-invariant monopole operators correspond to null space of the $\mb{K}$ matrix. Such operators only exist when the K matrix is degenerate, so it has zero eigenvalues, the number of which is the same as the dimension of the null space. We focus on one such mode, and denote by $b^0$. Since the CS term vanishes, the $b^0$ mode is just a pure Maxwell theory dual to a superfluid. The charged operator of the superfluid is the monopole operator.

A $2\pi m_I$ instanton for $a_I$ corresponds to a $2\pi e_I^q m_I$ instanton for the $b^q$ gauge fields. At this point, we pause to discuss the Dirac quantization condition. For $a^I$ gauge fields, we have the standard quantization condition:
\begin{equation}
    \int_{M_2} \frac{da_I}{2\pi} \in \Z,
\end{equation}
where $M_2$ is a closed surface. In terms of the new field strength:
\begin{equation}
    \sum_\alpha e^{q}_I\int_{M_2} \frac{f^q}{2\pi} \in \Z.
\end{equation}
Since we are only interested in gauge-invariant operators, we can assume that $\int_{M_2} f^q=0$ for $\lambda^q\neq 0$. For simplicity, we assume that $\mb{K}$ has a unique zero mode $\mb{v}$. The normalized eigenvector is $\mb{e}^0=\frac{\mb{v}}{\abs{\mb{v}}}$. Then the quantization condition is
\begin{equation}
    \frac{v_I}{\abs{\mb{v}}}\int_{M_2} \frac{f^0}{2\pi}\in \Z.
\end{equation}
Since gcd$(v_1, v_2,\dots)=1$, we find
\begin{equation}
    \int_{M_2} \frac{f^0}{2\pi \abs{\mb{v}}}\in\Z.
\end{equation}
This is the formal way of saying that the fundamental monopole is labeled by $\mb{v}$ (when expressed using the original gauge fields $a_I$).

To restore the standard normalization, we need to rescale $f^0\rightarrow \abs{\mb{v}}f^0$, so the Lagrangian density of the Maxwell term of $b^0$ now reads 
\begin{equation}
    \mathcal{L}_0=-\abs{\mb{v}}^2  \tilde{g}f^0\wedge \star f^0.
\end{equation}
This theory can be dualized to a superfluid, but now the superfluid density is proportional to $\frac{1}{\abs{\mb{v}}^2 \tilde{g}}$. So the mass generated by instanton proliferation is $\propto \frac{1}{|\mb{v}|^2}$ (there is a further fugacity factor, which should actually decay exponentially with $\abs{\mb{v}}^2$).
 
Now we specialize to the K matrix of the compactified system. In that case, we have $|\mb{v}|=\sqrt{N_z}$, thus the superfluid density goes down as $N_z^{-1}$. The monopole operator, which is charged under the magnetic symmetry of all the layers, is indeed the string superfluid order parameter of the coupled wire model identified in Sec. \ref{sec:stretched superfluid order}. The $N_z^{-1}$ decay of the superfluid density also agrees with the result of the two-point correlation function of the order parameter.

One notices an apparent discrepancy between the superfluid density which decays as $N_z^{-1}$ and a finite superconducting response. The resolution is that the charge of the order parameter grows with $N_z$ to compensate the decay of the superfluid density.

\subsection{Classification of charge excitations}

We now discuss the topological classification of charges in the Chern-Simons theory.
Again we consider the case of a finite number $N_z$ of layers.   Suppose the null space of the K matrix is spanned by null vectors $\mb{v}_j, j=1,2,\dots, r$ where $\mb{Kv}_j=0$. As shown earlier, each null vector corresponds to a 1-form $\U^{(j)}$ symmetry, and the corresponding magnetic U(1) 0-form symmetry is spontaneously broken. One can then show that the total vorticity (with respect to the weak superfluid of the $\U^{(j)}$ symmetry) is given by $\mb{v}_j^\mathsf{T}\mb{l}$, by e.g. checking the U(1)$^{(j)}$ 1-form charge of the corresponding Wilson loop. 

One can also consider the equivalence classes of non-vortex gauge charges.  The definition of equivalence is essentially the same as that of a topological Abelian CS theory with a non-degenerate K matrix, which we briefly review now: a gauge charge $\mb{l}$ is a local excitation if and only if it can be written as $\mb{l}=\mb{Kl}'$ for some integer vector $\mb{l}'$. This is true even when $\mb{K}$ is degenerate, as already mentioned in Sec. \ref{sec:monopole}, and easily follows from the equation of motion. Two charges are equivalent if their difference is local. From this definition, it is easy to see that all local excitations must have zero vorticity, as expected in any superfluid. We thus focus on the non-vortex charges, the equivalence classes of which form a finite Abelian group, which is the generalization of the anyon group. To determine the structure of this group, we can simply perform a $\mathrm{GL}(N_z, \Z)$ transformation to decouple the null space. In other words, we can find an invertible integer matrix $\mb{W}$, such that
\begin{equation}
    \mb{W}^\mathsf{T}  \mb{KW}=
    \begin{pmatrix}
    \mathbf{0}_{r\times r} & \\
    &  \tilde{\mb{K}}
    \end{pmatrix},
\end{equation}
where $\tilde{\mb{K}}$ is a non-degenerate matrix. The structure of the anyon group is determined by the Smith normal form of $\tilde{K}$. The number of distinct equivalence classes is given by $|\det \tilde{\mb{K}}|$. For the K matrix arising from the coupled wire construction, we find that $\tilde{\mb{K}}$ is basically the same as the K matrix with symmetry-breaking OBC, but with $N_z-1$ layers, and $|\det \tilde{\mb{K}}|=m^{N_z-1}\cdot N_z$. These results all agree very well that those of the coupled wire model.
 
\section{Summary and discussion}

In this work we have introduced a new type of compressible matter: a ``weak superfluid", which is a 3D anisotropic phase distinct from the conventional superfluid or Fermi liquid. The weak symmetry breaking is characterized by a ``rod" order parameter, which is supported on a straight line along a fixed direction, and at the same time there exist no local order parameters. We argue that the low-energy physics can be captured by a gapless infinite-component Chern-Simons theory introduced in Ref. [\onlinecite{ma2020fractonic}].

We determine the emergent symmetry of the highly anisotropic ``weak superfluid", both from the coupled wire model and from the iCS field theory. Most importantly, the weak symmetry breaking leads to a ``cylindrical" U(1) 1-form symmetry, where the symmetry transformations are defined on cylinders along the $z$ direction (with arbitrary shape in the $xy$ plane), as depicted in Fig \ref{fig:charge operators}. The mixed anomaly between the cylindrical U(1) 1-form symmetry and the U(1) 0-form symmetry is responsible for the arbitrary filling condition, i.e. being a compressible phase. 

Recently, Ref. [\onlinecite{ElseDrag2021}] identified an interesting conceptual relation between zero-temperature DC conductivity and ``fluxibility", which is a mixed anomaly between charge U(1) symmetry and an emergent symmetry. The latter is also responsible for the arbitrary filling condition. This led to the conjecture [\onlinecite{ElseDrag2021}] that a translation-invariant quantum system is compressible if and only if it is fluxible. In our work, the filling condition is indeed realized in the low-energy theory as a mixed anomaly between the U(1) charge conservation symmetry and the cylindrical 1-form symmetry, which suggests that the system is indeed ``fluxible", in an appropriate sense. However, the highly anisotropic emergent symmetry group results in more complicated zero-temperature DC transport, quite different from the isotropic theories considered in Ref. [\onlinecite{ElseDrag2021}].

Our microscopic model was given by a coupled wire construction which stitches together arrays of 1D Luttinger Liquids. We can also understand the phase by imagining gluing together layers of 2D superfluids. In  previous works [\onlinecite{FractonCW2021}] [\onlinecite{JoeCW2021}], other classes of 3D coupled wire models were shown to exhibit fractonic behavior. Interestingly, for appropriate choice of parameters, these models can also be made compressible (i.e. invariant under the anomalous $\U\times \U$ symmetry defined in Eq. \eqref{eqn:mixedU1}). Similar to the models studied in this work, they become gapless and exhibit weak symmetry breaking, i.e. there are no local order parameter of any kind. But the non-local order parameter is now supported on a membrane that spans the whole 2D section perpendicular to the wires. It is not yet clear whether the physics can be understood in terms of some variation of the iCS theory. We leave a detailed investigation of these other models for future work.

Another natural question is whether similar WSB phenomena occur in two-dimensional compressible systems. While we are not aware of any fundamental reasons prohibiting such phenomena in 2D, it is unlikely that a 2D coupled wire model can realize it. The duality transformation employed in this work can be applied to a general coupled wire model, under reasonable conditions. The result is then a 2+1D Abelian CS theory, which is either fully gapped when the K matrix is non-degenerate, or dual to a superfluid when the K matrix is degenerate, which now has a local order parameter (e.g. the quasi-2D limit of the model studied in this work after compactification). It will be interesting to understand whether this is just the limitation of the coupled wire construction, or there is a more fundamental obstruction in 2D.

The iCS models have been shown to realize new and novel phases of matter. The gapped case discussed in [\onlinecite{ma2020fractonic}] provides new examples of type-I fractonic order without any foliation structure.  Thus far the iCS theories that have been analyzed have only involved nearest neighbor couplings. It would be interesting to study models with a wider range of couplings between the layers of gauge theories,  and explore possible connections with other fracton phases. We have shown here that the coupled wire construction provides a useful tool for building microscopic models which realize these phases at low energy. 

\vspace{5mm}

\begin{acknowledgements}
M.C. would like to thank X. Chen for enlightening conversations and collaboration on a previous project, and D. Else and T. Senthil for useful correspondence and sharing unpublished results. J.S. would like to thank N. Read, X. Ma and S. Bryant for helpful discussions. J.S. especially thanks D. Mross and E. Leviatan for explaining subtleties in their recent work \cite{LeviatanPRR2020}. We are grateful to John McGreevy for his feedback on the manuscript. M.C. acknowledges support from NSF under award number DMR-1846109 and the Alfred P. Sloan
foundation. A.D. is supported by the Simons Foundation through the collaboration on Ultra-Quantum Matter (651438, XC) and by the Institute for Quantum Information and Matter, an NSF Physics Frontiers Center (PHY-1733907).

\emph{Note added:} We would like to draw the reader’s attention
to an upcoming work on closely related topics by X.-Q. Ma, H.-T. Lam and X. Chen~\cite{MaChenLam}. 
\end{acknowledgements}

\appendix
\onecolumngrid
\section{String operators for solitonic excitations}
\label{sec:Mobility of solitonic excitations}
In this section we consider the mobility of the solitons. We focus on the most elementary excitation, namely a  unit soliton of a single $\Theta_\mb{r}$ term. We work in the mean-field approximation, assuming large $vg$. 

 One can apply a vertex operator to create solitons, and move them around in the $yz$ plane. In this model, it is easy to see that $e^{i\theta}$ can hop a 1-soliton along $y$. In other words, $\prod_{y}e^{i\theta_{yz}(x)}$ is a string operator. Note that strictly speaking, the string creates perfectly sharp solitons, which are only valid at $g \rightarrow \infty$. At finite $g$, the soliton is smeared over a length scale $\xi$ and the string operator must be modified to create the smeared profile. But these modifications do not change universal features, such as mobility or the braiding statistics. 

We also consider how a single soliton excitation can be transported along the wire. {For simplicity we consider the $n_1=n_2=n$ case, but the same method works for the more general cases as well.} Following the general discussion in Ref. [\onlinecite{FractonCW2021}], we define 
\begin{equation}
    \varphi_{yz}^L = \varphi_{yz}+ m \theta_{yz}+n(\theta_{y,z-1}+\theta_{y,z+1}). 
\end{equation} 
    We construct a string operator of the following form: 
\begin{equation}
	W_y(x_2, x_1)=\exp \left( i \int_{x_1}^{x_2} \sum_z w_z\partial_x\varphi^L_{yz}\right).
	\label{eqn:Wx}
\end{equation}
where $w_z \in \mathbb{R}$. At the end point $x_2$, $W$ creates a soliton of strength $\sum_{z'}K_{zz'}w_{z'}$ for the $\Theta_{yz}$ term, where $y$ is fixed, and the opposite one at $x_1$. We thus denote such an excitation at one end of the string by a $N_z$-dimensional integer vector $\mb{v}$, the $z$-th entry of which is the strength of the soliton in the $\Theta_{yz}$ term. Thus finding a string operator of the form Eq. \eqref{eqn:Wx} reduces to solving the equation $\mb{Kw}=\mb{v}$.

Thus to move a single soliton of unit strength at $z=0$,  we set $w_z= (K^{-1})_{0z}$. 
When the bulk is fully gapped (i.e. $2|m|>|n_1+n_2|$), $w$ as a function of $z$ decays exponentially away from the location of the excitation (it is strictly localized in $y$), so our string operator is quasi-localized, and the soliton can move along the wire. This agrees with the prediction of the iCS field theory~\cite{ma2020fractonic}, that is the quasiparticles have exponentially localized profiles leading to quasi-localized braiding statistics. In the coupled wire model, the quasiparticle excitation itself is strictly localized as a violation of a certain plaquette term (at least in the mean-field limit), but the string operator is quasi-localized. 

In the gapless case, the construction in Eq. \eqref{eqn:Wx} does not work in general as $K$ is singular, at least when PBC along the $z$ direction is imposed.  However, we can still invert the K matrix in the complement of the null space. For $m=-2n$, this subspace consists of all vectors $\mb{v}$ such that $\sum_i v_i\neq 0$.  For the vectors with non-zero overlap with the null space, as we discussed in the main text they should be thought of as vortices in the weak superfluid. For example, if $\mb{v}_0=(1,-1,0,\cdots)$, then we can find the following $\mb{w}$ which satisfies $ \mb{Kw}=\mb{v}_0$:
\begin{equation}
    \mb{w}=\frac{1}{m}\Big(-\!\frac{1}{L_z}, \frac{L_z-2}{L_z},\frac{L_z-1}{L_z},\cdots,\frac{1}{L_z},0\Big).
\end{equation}
$\mb{w}$ is no longer quasi-localized. The weight is only inversely proportional to the distance away from the location of the excitation. A string operator moving this excitation along the wire direction can then be constructed using the $\mb{w}$.

 It is shown in Ref. [\onlinecite{JoeCW2021}] that two excitations $\mb{v}$ and $\mb{v}'$ (with the same $x$ coordinate) belong to the same superselection sector, i.e. they can be transformed into each other by acting with local operators, if and only if $\mb{v}'=\mb{v}+\mb{Ku}$ for some integer vector $\mb{u}$. This agrees with the mathematical definition of superselection sectors in Abelian CS theories. Let us now enumerate the number of superselection sectors. We can easily show that, all translations of $m\mb{v}_0$ are equivalent, and $mL_z\mb{v}_0$ is local.  It is then straightforward to show that the total number of inequivalent non-vortex excitations is $m^{L_z-1}\cdot L_z$. More generally, we can find the Smith normal form of $\mb{K}$, and the number of superselection sectors is the same as the absolute value of the product of all non-zero entries. Notice that here we only count the ``non-vortex" excitations, as there are infinite types of vortices labeled by vorticity. 

Finally, we consider the mobility of excitations along $z$. Since $e^{i\theta}$ only moves solitons along $y$, motion along $z$ necessarily involves $e^{i\varphi}$.   It is useful to represent the charge configuration formally by a Laurent polynomial in two variables $y$ and $z$. Namely, define
\begin{equation}
    \sum_{i,j\in \Z}q_{ij}y^i z^j.
\end{equation}
Here $q_{ij}$ is the charge at coordinate $i,j$ in the $y-z$ plane and it should be clear that we do not need to worry about their $x$ coordinates. Such polynomial representation was discussed in Ref. [\onlinecite{FractonCW2021}].

The operator $e^{i\varphi}$ creates the pattern $(nz^{-1}+m+nz)(1+y)$. Since $1-y$ can be created by $e^{i\theta}$, for mobility along $z$ we can move all excitations to the same $y$ and thus just consider $2(nz^{-1}+m+nz)$. Suppose by applying $e^{ia_j\varphi_j}$ (together with appropriate $e^{i\theta}$'s to move all charges to the same $y$), we can create a configuration with two solitons of opposite charges separated by a distance $l$. In terms of the polynomial representation, this amounts to finding a polynomial $f(z)=\sum_j a_j z^j$ such that 
\begin{equation}
    f(z)(nz^{-1}+m+nz)\propto 1+z^l.
\end{equation} 
Suppose that $l$ is large so we essentially look for a string operator along $z$. For $1\ll j\ll l$ we have
\begin{equation}
	na_{j+1}+ma_j+na_{j-1}=0.
	\label{}
\end{equation}
Since $m^2>4n^2$, the corresponding characteristic polynomial has two real roots. As a result, $a_j$ grows exponentially with $j$ (in either directions), which means the charge created at one end costs an exponentially large amount of energy. For $m=-2n$,  let $f(z)=a_0+a_1z+\cdots+a_l z^l$, then $f(z)(nz^{-1}-2n+nz)=n[a_0+(a_1-2a_0)z+(a_2-2a_1+a_0)z^2+\cdots  + a_l z^{l+1}]$. If we require $a_{j+1}-2a_j+a_{j-1}=0$ for all $1\leq j\leq l-1$, then we $a_j=a_0+j(a_1-a_0)$. So the only way to have bounded $|a_j|$ is to set $a_1=a_0$, and the excitation created at one end is precisely a vortex-anti-vortex dipole of strength $n$. 

\section{Details of the duality mapping
\label{sec:Details of Duality Mapping}}
In this section we provide more detail for the mapping described in section \ref{sec:Duality mapping}.
\subsection{Boson-vortex duality}
The wire model is described by an array of Luttinger liquids with conjugate variables $(\varphi, \theta)$ and Lagrangian
\begin{equation} 
\label{eq:WireLagrangian1 app} 
\begin{split}
\mathcal{L}[\varphi,\theta]&=\sum_\vr \frac{i}{\pi}\partial_x \theta_\vr \partial_\tau \varphi_\vr+\frac{\tilde{v}}{2 \pi}\left(\partial_{x} \varphi_{\vr}\right)^{2}+\frac{u}{2 \pi}\left(\partial_{x} \theta_{\vr}\right)^{2}+\frac{v}{8 \pi}\left(\partial_{x} \Delta_y \varphi_{\vr}\right)^{2} - g\cos(2\bth_\vr)  
\end{split}
\end{equation} 
where 
\begin{equation}
\label{eq:Plaquette def app}
    \begin{split}
        2\bth_{\vr} &= \Delta_y \varphi_\vr + \Lambda \theta_\vr\\ 
        &=\left(\varphi_{\vr+\hat{y}}-\varphi_{\vr}\right)+ \left(m \theta_{\vr} +m\theta_{\vr+\hat{y}} +n_1\theta_{\vr-\hat{z}}+n_2\theta_{\vr+\hat{z}}+n_2\theta_{\vr+\hat{y}-\hat{z}}+n_1\theta_{\vr+\hat{y}+\hat{z}}\right). \end{split}
\end{equation}
Here we define
 $\Delta_y X_\vr = X_{\vr+\hat{y}}-X_\vr$ and $\vr=(y,z)$ is a wire index. The definition of $\Lambda$ can be easily inferred from Eq. \eqref{eq:Plaquette def app}. Note also that $\frac{v}{8 \pi}\left(\partial_{x} \Delta_y \varphi_{\vr}\right)^{2}$ has been added to the standard Luttinger Liquid kinetic term. This is done because it is convenient for the duality mapping to the vortex theory but its presence does not affect the qualitative physics \cite{mrossPRX2017}.
 
 In what follows we use the two equivalent forms of labelling for the wires, $O_{\vr+\hat{a}} \equiv O_{y+a_y,z+a_z}$. Anticipating the layered structure of the gauge theory, we treat $z$ as the ``layer" index and $y$ is to be coarse grained to a continuous spatial coordinate. With this motivation in mind, 
 we define the following pair of conjugate variables: 
\begin{equation}
\label{eq:vortexvar app}
\begin{array}{l}
\tph_{yz}=-\sum_{y^{\prime}} \sgn\left(y^{\prime}-y-\frac{1}{2}\right) \theta_{y^{\prime}z} \\
\\
\tth_{yz}=\frac12\left(\varphi_{y+1,z}-\varphi_{y,z}\right) . 
\end{array}
\end{equation}
Here $\tph_{yz}$ creates a $2\pi$ vortex in the $\varphi$ field in layer $z$, in between wires $y$ and $y+1$. The operator $\partial_x \tth_{yz}$ is the ``charge" operator for this vortex.
We can re-express the Lagrangian in Eq. \eqref{eq:WireLagrangian1 app} in terms of these new fields but the result will be  highly non-local in the $y$ direction. We can restore locality via a Hubbard-Stratonovich transformation at the expense of introducing new degrees of freedom $a_{0}^{(z)}(x,y)$ and $a_{1}^{(z)}(x,y)$ in each layer indexed by $(z)$.

We now rewrite the kinetic part of the Lagrangian using the dual variables $(\tph,\tth)$ and the Hubbard-Stratonovich fields $(a_0^{(z)},a_1^{(z)})$:
\begin{equation}
\begin{array}{l}
\sum_\vr \frac{i}{\pi}\partial_x \theta_\vr \partial_\tau \varphi_\vr = \sum_\vr \frac{i}{\pi}\partial_x \tth_\vr \partial_\tau \tph_\vr \\

  \\ \sum_\vr\frac{v}{8 \pi}\left(\partial_{x} \Delta_y \varphi_{\vr}\right)^{2} = \sum_\vr \frac{v}{2 \pi}\left(\partial_{x} \tth_{\vr}\right)^{2}
    \\
    
\\
    \sum_\vr\frac{\tilde{v}}{2 \pi}\left(\partial_{x} \varphi_{\vr}\right)^{2} = \sum_\vr \frac{\tilde{v}}{2 \pi}\left(\Delta_y^{-1}\partial_{x} \tth_\vr\right)^{2} \to \sum_\vr \left[ -\frac{i}{\pi}\partial_x \tth_\vr a_{0}^{(z)} + \frac{(\Delta_y a_{0}^{(z)})^2}{8\pi \tilde{v}} \right]
    \\
    \\
    \sum_\vr \frac{u}{2\pi}\left(\partial_{x} \theta_{\vr}\right)^{2} = \sum_\vr \frac{u}{8 \pi}\left(\partial_{x} \Delta_y \tilde{\varphi}_{\vr}\right)^{2} \rightarrow  \sum_\vr \frac{u}{8 \pi}\left[\left(\partial_{x} \tilde{\varphi}_{\vr}-a_1^{(z)}\right)^{2}+\left(\Delta_y a_{1}^{(z)}\right)^{2}\right]
+\sum_{z,y, y^{\prime}} \frac{u}{8 \pi} V_{y, y^{\prime}} \partial_{x}\left(\Delta_y \tilde{\varphi}_{yz}\right) \partial_{x}\left(\Delta_y \tilde{\varphi}_{y^{\prime}z}\right)
\end{array}
\end{equation}
where $V = \Delta_y^\mathsf{T} (1+\Delta_y^\mathsf{T} \Delta_y)^{-1} \Delta_y$, which decays exponentially in the difference in wire index $y$. Up to this point everything is exact. Now we interpret $a_\mu^{(z)}$ as a (2+1)d gauge field, living in layer $z$, in the $a_2^{(z)}=0$ gauge. The upshot of all this is that we get a theory of layers of vortices minimally coupled to a layered gauge theory. Using Eq. \eqref{eq:vortexvar app}, we re-express the plaquette term in terms of the vortex variables
\begin{equation}
\label{eq:Plaquette vortex def app}
    \begin{split}
        2\bth_{yz}
	&= 2\tth_{yz} + \frac12\Delta_y \left[n_1 (\tph_{y, z+1} +\tph_{y+1,z-1}) + n_2 (\tph_{y,z-1} +\tph_{y+1,z+1}) + m (\tph_{yz} +\tph_{y+1,z})\right]\\
	&= 2\tth_{yz} + \frac{1}{2}\Lambda\cdot \Delta_y \tph_{yz} ~~.
    \end{split}
\end{equation}
In what follows, when coupled to matter fields $(\bphi,\bth)$, we suppress the $(z)$ layer index of the gauge fields and use the $\vr$ index in order to make the expressions more succinct. The various indices can be parsed as follows $a_{\mu,\vr} = a_{\mu, yz} = a_\mu^{(z)}(x,y)$. At this stage the Lagrangian takes the form
\begin{equation}
\label{eq:Vortex_Lag app}
\mathcal{L}[\tph,\tth,a_\mu] = \sum_\vr \frac{i}{\pi} \partial_{x} \tilde{\theta}_\vr \left(\partial_\tau \tph_\vr -a_{0,\vr}^{(z)}\right) + \frac{u}{2 \pi}\left(\partial_{x} \tph_{\vr}-a_{1,\vr}^{(z)}\right)^{2}+\frac{v}{2 \pi}\left(\partial_{x} \tth_\vr\right)^{2} - g\cos(2\tth_\vr +\frac{1}{2}\Lambda \cdot \Delta_y \tph_\vr) + \mathcal{L}_{\text{Maxwell}}^{(z)}.
\end{equation}

Here, in the $a_{2}^{(z)}=0$ gauge, $\mathcal{L}_{\text{Maxwell}}^{(z)} = \frac{u}{2\pi}(\Delta_y a_1^{(z)})^2 + \frac{1}{2\pi \tilde{v}}(\Delta_y a_0^{(z)})^2$. In order to simplify the sine-Gordon term we can introduce another set of conjugate variables:
\begin{equation}
\label{eq:Plaq variables app}
  \bphi_{\vr} = \tph_{\vr} ~~~ \text{and} ~~~ \bth_\vr = \tth_\vr +\frac{1}{4}\Lambda\cdot \Delta_y \bphi_\vr.
\end{equation}
Expressing the Lagrangian in this final basis and also expanding the cosine term yields
\begin{equation}
\label{eq:Plaquette_Lag app}
\begin{split}
\mathcal{L}[\bphi,\bth,a_\mu] = &\sum_\vr\frac{i}{\pi} \partial_{x} \bth_\vr (\partial_\tau \bphi_\vr -a_{0, \vr}) + \frac{i}{4\pi} \left(\Lambda \cdot \Delta_y \partial_x \bphi\right)_\vr a_{0,\vr}
+\frac{u}{2 \pi}\left(\partial_{x} \bphi_{\vr}-a_{1, \vr}\right)^{2}\\
&~~~+\frac{v}{2 \pi}\left(\partial_{x} \bth_\vr\right)^{2}
-\frac{v}{4 \pi}\left(\Lambda \cdot \Delta_y \partial_x \bphi \right)_\vr \partial_x\bth_\vr + \frac{v}{32 \pi}\left(\Lambda \cdot \Delta_y \partial_x \bphi \right)_\vr^2\\
&~~~+g\bth_\vr^2 + \mathcal{L}_{\text{Maxwell}} .
\end{split}
\end{equation}

\subsection{Integrating out matter fields}
In order to get a pure gauge theory the plaquette degrees of freedom $\bphi, \bth$ must be integrated out. 
The first step in the derivation is to collect all the terms involving $\bphi$, {up to $\mathcal{O}(\bphi^2)$}:
\begin{equation}
    \begin{split}
        &\mathcal{O}(\bphi^2): ~  \partial_x \bphi_{\vr^\prime}\left[\frac{u}{2\pi} + \frac{v}{32\pi} (\Lambda \cdot \Delta_y)^\mathsf{T}(\Lambda \cdot \Delta_y)\right]_{\vr^\prime \vr}\partial_x \bphi_\vr = \partial_x \bphi M \partial_x \bphi\\
        &\mathcal{O}(\bphi): ~ \underbrace{\left[\frac{-i}{4\pi}\Lambda \cdot \Delta_y a_{0} - \frac{u}{\pi}a_{1} + \frac{i}{\pi}\partial_\tau\bth + \frac{v}{4\pi}\Lambda \cdot \Delta_y \partial_x \bth \right]_\vr}_{\Gamma_\vr} \partial_x\bphi_\vr
    \end{split}
\end{equation}
We denote the cross term by $\Gamma_\vr[\bth,a] \partial_x\bphi_\vr .$
Note here we have used a discrete ``integration by parts" in parts involving $\Lambda\cdot \Delta_y$ and dropped $z$ boundary terms. This step requires care when considering open boundary conditions in $z$. Integrating out $\bphi$ results in an expression of the form:
\begin{equation}
    -\frac{1}{4} \Gamma_{\vr^\prime} \left(M^{-1}\right)_{\vr^\prime \vr} \Gamma_\vr.
\end{equation}
 We are interested in the low energy theory so we can express $M^{-1}$ as a derivative expansion in powers of $\Lambda \cdot \Delta_y$. To accomplish this note that for $M= 1+D$
\begin{equation}
\begin{split}
    M^{-1} &= \left(1+D\right)^{-1} 
    = \sum_{j=0}^\infty (-1)^j  D^j~.
\end{split}
    \end{equation}
    In the present case $M^{-1} \approx \frac{2\pi}{u} - \frac{v\pi}{8u^2}(\Lambda\cdot \Delta_y)^\mathsf{T}(\Lambda\cdot \Delta_y)$.
    We keep the 2nd order term because the $a_1^2$ term in $\Gamma^2$ cancels and the next lowest order term is $(\Lambda\cdot\Delta_y a_1)^2$. Including the left over terms which do not involve $\bphi$ and only keeping the lowest order derivative terms in the expansion of $\Gamma^2$ gives
  \begin{equation}
  \label{eq:L theta and a app}
      \begin{split}
         \mathcal{L}[\bth, a] &= \sum_\vr \frac{-i}{4\pi} (\Lambda\cdot \Delta_y a_0)_\vr a_{1,\vr} +\frac{v}{8\pi} (\Lambda\cdot \Delta_y a_1)_\vr^2
         + \frac{1}{32\pi u } (\Lambda\cdot \Delta_y a_0)_\vr^2 +  \frac{i}{\pi}  \bth_\vr \left(\partial_x  a_0 -\partial_\tau a_1 \right)_\vr + g \bth_\vr^2\\
         &~~~~+ \mathcal{L}_{\text{Maxwell}} + \text{Higher order terms}
      \end{split}
  \end{equation}
Integrating out $\bth$ is now relatively simple and the result is
\begin{equation}
\label{eq:L a app}
    \begin{split}
        \mathcal{L}[a] = &\sum_\vr -\frac{i}{4\pi}  \underbrace{(\Lambda\cdot \Delta_y a_1)_\vr a_{0,\vr}}_\text{CS term} + \frac{1}{4\pi^2g}(\partial_x a_0 - \partial_\tau a_1)_\vr^2 \\
        &~~~~~~~+\frac{1}{2\pi \tilde{v}}(\Delta_y a_0)_\vr^2 +\frac{1}{32\pi u} (\Lambda\cdot \Delta_y a_0)_\vr^2
         + \frac{u}{2\pi}(\Delta_y a_1)_\vr^2 + \frac{v}{8\pi} (\Lambda\cdot \Delta_y a_1)_\vr^2 .
    \end{split}
\end{equation}

 If the $y$ direction is coarse grained and the layer $(z)$ index restored then
\begin{equation}
\begin{split}
 \left(\Lambda \cdot \Delta_y a\right)_z &\approx (n_1 + n_2)\Delta_y a^{(z+1)} + 2m \Delta_y a^{(z)} + (n_1 + n_2) \Delta_y a^{(z-1)}
 \end{split}
\end{equation}

So the first term in Eq. \eqref{eq:L a app}, corresponding to a CS term, becomes
\begin{equation}
    \mathcal{L}_\text{CS}[a] = \frac{i}{4\pi}\left[(n_1+n_2) a_1^{(z+1)} \partial_y a_0^{(z)} + 2m a_1^{(z)} \partial_y a_0^{(z)} + (n_1+n_2) a_1^{(z-1)} \partial_y a_0^{(z)}\right] ~~~.
\end{equation}
When the gauge constraint, $a_2=0$, is relaxed $a^{(z)}_1 \partial_y a_0^{(z^\prime)} \to \epsilon^{\mu\nu\lambda} a^{(z)}_\mu \partial_\nu a_\lambda^{(z^\prime)}$, so the CS part of the Lagrangian has the more familiar form
\begin{equation}
    \mathcal{L}_\mathrm{CS}[a] \equiv \frac{iK_{zz^\prime}}{4\pi} a^{(z)} \wedge da^{(z^\prime)} ~~.
\end{equation}

We can similarly investigate parts of Eq. \eqref{eq:L a app} associated with the Maxwell terms of the gauge theory. For example, coarse graining $y$ and restoring gauge invariance in the $4^{\text{th}}$ term corresponds to
 \begin{equation}
 \label{eq:E2 app}
    (\Lambda\cdot \Delta_y a_0)^2 \to (\Lambda\cdot [ \Delta_y a_0-\partial_\tau a_2])^2 = \left[(n_1 + n_2) E_2^{(z+1)} + 2m E_2^{(z)} + (n_1 + n_2) E_2^{(z-1)}\right]^2 ~~.
\end{equation}
We see that the typical $E_y^2$ term present in conventional Maxwell theory is replaced by something which couples the same component ($y$ in this case) of the electric field in different layers. Indeed, a similar term is generated for $B$ but not for $E_x$, this is not surprising given the anisotropic nature of the underlying microscopic model. The full result for the Maxwell term is given by
\begin{equation}
\label{eq:Layered Spectrum}
    \mathcal{L}_{\text{Maxwell}} = \sum_{zz^\prime} E_1^{(z)} \frac{\delta_{zz^\prime}}{4\pi^2g} E_1^{(z^\prime)} + E_2^{(z)}\left[ \frac{\delta_{zz^\prime}}{2\pi\tilde{v}} + \frac{1}{32\pi u}(K^\mathsf{T}K)_{zz^\prime} \right] E_2^{(z^\prime)} + B^{(z)}\left[ \frac{u \delta_{zz^\prime}}{2\pi} + \frac{v}{8\pi}(K^\mathsf{T}K)_{zz^\prime} \right] B^{(z^\prime)} ~~.
\end{equation}
The upshot of this discussion is that the effective IR Lagrangian is given by:
\begin{equation}
    \mathcal{L}[a] \equiv \frac{iK_{zz^\prime}}{4\pi} a^{(z)} \wedge da^{(z^\prime)} + \mathcal{L}_{\text{Maxwell}} ~~.
\end{equation}

Note that for the compressible state, $m=-(n_1+n_2)$, $K^\mathsf{T}K\to \partial_z^4$ under coarse-graining in $z$ and so the inter-layer portion of the Maxwell term would appear to be irrelevant in this context.

\subsection{Coupling to vortices}
\label{sec:Coupling to vortices}
So far we have assumed that $\varphi$ and $\theta$ vary smoothly, excluding the ``solitonic" excitations of the cosine pinning term. These more singular configurations are the analog of vortices in the usual boson-vortex duality, which should be minimally coupled to the dynamical gauge fields. we now extend the duality map to include these excitations.

To this end, we replace the cosine potential $g\cos (2\bth_\vr)$ by 
\begin{equation}
    \frac{-g}{2}(2\bth_\vr-2\pi n_\vr)^2
\end{equation}
Here $n_\vr$ parametrizes the locations of the solitons where $2\bth$ jumps by integer multiples of $2\pi$.

From the discussion of the microscopic model we can view $n$ as corresponding to some configuration of vortices created by an operator of the form $e^{i\sum\int n(x,y,z)\partial_x \tph}$.  It is natural to wonder how the introduction of such a field manifests in the dual gauge theory. This can be settled using the mapping of section \ref{sec:Duality mapping}. Indeed, amending Eq \ref{eq:L theta and a app} gives
 \begin{equation}
  \label{eq:L theta and a and n}
      \begin{split}
         \mathcal{L}[\bth, a] &= \sum_\vr \frac{-i}{4\pi} (\Lambda\cdot \Delta_y a_0)_\vr a_{1,\vr} +\frac{v}{8\pi} (\Lambda\cdot \Delta_y a_1)_\vr^2 + \frac{1}{32\pi u } (\Lambda\cdot \Delta_y a_0)_\vr^2\\
         &~~~~+ \frac{i}{\pi}  \bth (\partial_x  a_0 -\partial_\tau a_1  + 2\pi i g n) + g \bth_\vr^2 + \frac{g}{4}n_\vr^2\\
         &~~~~+ \mathcal{L}_{\text{Maxwell}} + \text{Higher order terms}.
      \end{split}
  \end{equation}
  Integrating out $\bth$ gives Eq. \eqref{eq:L a app} plus a coupling term between the gauge fields and $n$
  \begin{equation}
      \label{eq:L a and n}
      \mathcal{L}[a,n] = \mathcal{L}[a] + i(\partial_x a_0^{(z)} - \partial_\tau a_1^{(z)}) n \rightarrow \mathcal{L}[a] - (a_0  i\partial_x n - a_1 i\partial_\tau n)\\.
      \end{equation}
  This tells us that we can interpret $n$ as a $x$-dipole of the gauge charge. Namely, $\partial_x n = \rho$ is the charge density. One can expect that if we treat the dynamics of solitons carefully (e.g. solitons moving between wires), one would obtain the full matter-gauge coupling $j\cdot a$.  Therefore, as expected, vortices become gauge charges.

\section{Quantization condition for 1-form symmetry transformation}
\label{sec:quantization}

In order to derive the quantization conditions on $q_I$, it is more convenient to use the 4d definition of the Chern-Simons term. Suppose that $M_3$ is the boundary of a 4d manifold $M_4$, i.e. $\partial M_4=M_3$. The gauge field is also extended to $M_4$. Then the CS action can be defined as
\begin{equation}
    S=\frac{K_{IJ}}{4\pi}\int_{M_4}F_I\wedge F_J.
\end{equation}
Now consider 1-form transformations parametrized by $\lambda_I$. We assume that they are extended to the bulk as well, but not necessarily flat. The 4d action changes:
\begin{equation}
\begin{split}
    \delta S 
    &= \frac{K_{IJ}}{4\pi}\int_{M_4}(  q_I d\lambda_I\wedge F_J + q_J F_I\wedge d\lambda_J + q_I q_J d\lambda_I \wedge d\lambda_J)\\
    &=\frac{K_{IJ}}{4\pi}\int_{M_4}( q_I d\lambda_I\wedge F_J + q_J F_I\wedge d\lambda_J)  +\frac{K_{IJ}}{4\pi}q_Iq_J\int_{M_3}\lambda_I \wedge d\lambda_J\\ &=\frac{K_{IJ}}{4\pi}\int_{M_4}( q_I d\lambda_I\wedge F_J + q_J F_I\wedge d\lambda_J)\\ 
    &=2\pi K_{IJ} q_I \int_{M_3}\frac{\lambda_I}{2\pi}\wedge \frac{F_J}{2\pi}
\end{split}
\end{equation}
Since $\int_{M_3}\frac{\lambda_I}{2\pi}\wedge \frac{F_J}{2\pi}$ is an integer (intersection number), for the partition function to remain the same, we must have $e^{i\delta S}=1$, or $K_{IJ}q_I$ an integer. 


\section{Phenomenologies}
\subsection{Order parameter}
\label{sec:order_parameter}
 The stretched superfluid order parameter is given by the string operator $e^{i\Phi_y}$ where $\Phi_y(x) = \sum_z \varphi_{yz}$. We want to compute the two point function:
$\langle e^{i\Phi_y(x)} e^{-i\Phi_{y^\p}(x^\p)} \rangle
$. To tackle this we approximate the sine-Gordon term by the quadratic term $\Theta^2$ in the strong coupling limit. The resulting theory is free so we can shift our attention to calculating the $\langle \Phi_y(x)\Phi_{y^\p}(x^\p) \rangle$, which is equal to
\begin{equation}
\begin{split}
    \sum_{z,z^\prime} \langle \varphi_{yz}(x)\varphi_{y^\prime z^\prime}(x^\prime)\rangle &= \SumInt_q \SumInt_{q^\prime}\langle \varphi_q \varphi_{q^\prime} \rangle \sum_{z,z^\prime}e^{iq\cdot r}e^{iq^\prime \cdot r^\prime }
    = \SumInt G_\varphi(q) e^{ik_\perp\cdot(r -r^\prime)}\sum_{z, z^\prime=1}^{N_z} e^{ik_z(z-z^\prime)} .
    \end{split}
\end{equation}
Here $k_\perp=(\omega,k_x,k_y)$ is the momentum perpendicular to $z$, $q = (k_\perp,k_z)$ and $G(q,q^\p)=\delta_{q,q^\p}G(q)$, since the theory is free and translation invariant. The sum over $z,z'$ can be easily evaluated
\begin{equation}
    \sum_{z z^\p} e^{ik_z (z-z^\p)} = N_z \sum_z e^{ik_z z} = N_z\delta(k_z).
\end{equation}
Thus 
\begin{equation}
\begin{split}
    \langle \Phi_y(x)\Phi_{y^\p}(x^\p) \rangle &= N_z \int dk_\perp\, e^{i k_\perp \cdot \vr} G_\varphi(k_z=0).
    \end{split}
\end{equation}

To work out the two-point function $G_\varphi$ we re-express the action as 
\begin{equation}
\label{eq:Meanfield_S}
S = \sum_q 
\begin{pmatrix}
\varphi(q) & \theta(q)
\end{pmatrix}
G^{-1}(q)
\begin{pmatrix}
\varphi(-q) \\
\theta(-q)
\end{pmatrix} ~.
\end{equation}
Recall the mean-field {Lagrangian} Eq. \eqref{eq:WireLagrangian1} is given by:
\begin{equation}
    S = \SumInt_{\vr,x,\tau}dxd\tau \frac{i}{\pi}\partial_x \theta_\vr \partial_\tau \varphi_\vr + \left[\tilde{v} (\partial_x \varphi_\vr)^2 + u (\partial_x \theta_\vr)^2  + g \Theta_\vr^2 \right] ~~.
\end{equation}

{Using this, we get} 
\begin{equation}
    G^{-1}(q) = \begin{pmatrix}
\tilde{v} k_x^2 + gf_\varphi(q) & \frac{ik_x \omega}{\pi} + g f(q)\\
 &\\
\frac{ik_x \omega}{\pi} + g f(-q) & u k_x^2 + g f_\theta(q)
\end{pmatrix}
\end{equation}

where
\begin{equation}
\begin{array}{l}
f_\varphi = 2-2\cos k_y\\

\\

f_\theta = 4\left[m\cos\frac{k_y}{2}+n_1 \cos (\frac{k_y}{2}-k_z) + n_2 \cos (\frac{k_y}{2}+k_z)\right]^2
\\
\\

f =
2i\big[n_1 \sin k_z + n_2 \sin (k_y+k_z) + m \sin k_y+ n_1 \sin \ (k_y-k_z) - n_2 \sin k_z\big] ~~~.\\ 
\end{array}
\end{equation}

We are interested in realizations with the chiral U(1) present which occurs when $n_1+n_2=-m$. In this case $f_\varphi f_\theta - |f|^2=0$ and so we can express $G_\varphi$ as follows
\begin{equation}
    G_\varphi = \frac{u k_x^2 + gf_\theta}{k_x^2\left[\tilde{v} u k_x^2 + \frac{\omega^2}{\pi^2} + g\left(u f_\varphi + \tilde{v} f_\theta \right) \right] }.
\end{equation}
Setting $k_z=0$, we have $f_\theta=0$, and $G_\varphi$ is simplified to
\begin{equation}
    G_\varphi = \frac{u}{ \frac{\omega^2}{\pi^2} +u\tilde{v} k_x^2 +  2gu(1-\cos k_y) }.
\end{equation}
This form makes sense as $k_z=0$ means $\theta$ fields are independent of $z$, so the wire coupling term decouples into $\cos(\varphi_{yz}-\varphi_{y+1,z})$. For long wavelength limit, we can expand $1-\cos k_y\approx \frac{k_y^2}{2}$, and rescale $k_x\rightarrow \frac{k_x}{\sqrt{\tilde{v}}}, k_y\rightarrow \frac{k_y}{g}$,
\begin{equation}
    \langle \Phi_y(x)\Phi_{0}(0) \rangle \sim \frac{N_z}{\sqrt{gx^2+\tilde{v} y^2}},
\end{equation}
which can be understood as a superfluid with anisotropic dispersion.
Thus the two-point function for our order parameter has the following scaling:
\begin{equation}
    \langle e^{i\Phi_y(x)}  e^{-i\Phi_0(0)}\rangle \sim e^{-C\frac{N_z}{\sqrt{gx^2+\tilde{v}y^2}}}.
\end{equation}
Here $C$ is a constant.

\subsection{Compressibility}
\label{sec:Compressibility}
The presence of both global U(1) symmetries means the model can be defined at arbitrary filling. This, in turn, suggests compressibility. Here we add more field theoretic justification for this feature. To compute compressibility we add a uniform chemical potential term to the Hamiltonian
\begin{equation}
\label{eq:uniform mu}
    \begin{split}
        H-\mu N &= H - \mu \sum_\vr\int \partial_x \theta_\vr\\
        &= \int \frac{1}{2\pi}[\tilde{v}(\partial_x \varphi)^2 + u(\partial_x \theta)^2] - \mu \partial_x \theta - g \cos(2\bth)
    \end{split}
\end{equation}
where we have suppressed $\sum_\vr$ in the last line. The term $\mu \partial_x \theta$ can be absorbed into $(\partial_x\theta)^2$ by completing the square: $\frac{u}{2\pi}(\partial_x \theta)^2 - \mu \partial_x \theta = \frac{u}{2\pi}(\partial_x \theta-\frac{\pi}{u}\mu)^2 -\frac{\pi }{2u} \mu^2$. Next we can shift $\partial_x \theta-\frac{\pi}{u}\mu \to \partial_x \theta$ by redefining $\theta \to \theta + \frac{\pi}{u}\mu x$ but this is also a symmetry of the sine-Gordon term. In particular under this transformation $2\bth \to 2\bth$. Including the uniform chemical potential the action is given by
\begin{equation}
    S[\mu]= S- \frac{\pi V}{2u} \mu^2. 
\end{equation}
The compressibility is $\kappa \sim \frac{\partial^2 F}{\partial \mu^2}$ which, for this action, is finite. Note that the above argument rests on the fact that the shift $\theta \to \theta + \frac{\pi }{u}\mu x$ is a symmetry of $2\bth$. This is a consequence of the presence of the $U(1)_\theta$ subsystem symmetry. So we are in some way factoring in the stretched order parameter in this calculation.


\subsection{Electromagnetic response}
\label{sec:Electromagnetic response}
In order to work out the transport properties of the model we can consider coupling the gauge theory to a 3+1D background field $A$. The natural current is the monopole current in each layer $j_\mu^I \sim \epsilon^{\mu \nu \lambda}\partial_\nu a^{I}_\lambda$. Indeed the appropriate minimal coupling can be derived from the microscopic model by replacing $\partial_\mu \varphi \to \partial_\mu \varphi + A_\mu$ and carrying out the mapping of section \ref{sec:Duality mapping}. For simplicity consider the case with an intra-layer Maxwell term and a common charge $t$ in each layer as in section \ref{sec:monopole},
\begin{equation}
\label{eq:Transport Lag 1}
     \mathcal{L}=\frac{K^{IJ}}{4\pi} a_I \wedge da_J -  \tilde{g} f_I \wedge \star f_I + \frac{it}{2\pi} A_I \wedge da_I
\end{equation}
where here $A_I = A(x,y;z=I)$. Note the $A_z(x,y,z)$ does not couple to any current, which reflects the fact that the transport of charge along layers is highly suppressed. This can be seen from the form of the coupling term in the microscopic model. Recall the sine-Gordon term $\Theta_{yz} = \varphi_{yz} - \varphi_{y+1,z} + (\theta \text{ terms })$. Heuristically this condenses a process in which a charged particle created by $e^{i\varphi}$ is hopped along the y direction with some additional complicated back-scattering of vortices between different $z$ layers, but no charges tunnel between layers. As was done in Sec. \ref{sec:monopole}, we can diagonalize Eq. \eqref{eq:Transport Lag 1}. 
\begin{equation}
     \mathcal{L}=\sum_q\frac{\lambda^q}{4\pi} b^q \wedge db^q -  \tilde{g} f^q \wedge \star f^q + \frac{it}{2\pi} A^q \wedge d b^q.
\end{equation}
In what follows we use $q$ subscripts and superscripts interchangeably, (e.g. $O_q \equiv O^q$) since there is no subtlety with raising and lowering operators for the layer index. 

It is helpful to integrate out the gauge fields $b$ in order to get a theory purely in terms of the background field $A$. Note that $\mathcal{L} = \sum_q \mathcal{L}_q$ where, defining $m_q = \frac{\lambda_q}{8\pi q}$ and working in the $\partial_\mu b^q_\mu=0$ gauge,
\begin{equation}
    \begin{split}
        \mathcal{L}_q &= - \tilde{g}f^{\mu \nu, q} f_{\mu\nu}^q + 2m_q \epsilon^{mu \nu \gamma} b_\mu^q \partial_\nu b^q_\gamma + b_\mu^q J^{\mu, q}
        = 2  \tilde{g} ~ b^q_\mu \left( \partial^2 \eta^{\mu\nu} - m_q \epsilon^{\gamma \mu \nu}\partial_\gamma \right) b^q_\nu + b_\mu^q J^{\mu, q} ~~~.
    \end{split}
\end{equation}
One can check that the propagator for this theory is given by
\begin{equation}
\label{eq:propagator}
    \begin{split}
        D^{\mu\nu}_{qq^\prime}(x,y) = \frac{-\delta_{qq^\prime}}{2 \tilde{g}}\int \frac{d^3k}{(2\pi)^3} \frac{1}{k^2 - m_q^2} \left[\eta^{\mu\nu} -m_q^2\frac{k^\mu k^\nu}{k^4} + i \frac{m_q}{k^2} \epsilon^{\mu\nu\gamma}k_\gamma \right]e^{ik\cdot (x-y)} ~~.
    \end{split}
\end{equation}
Integrating out $b_q$ results in the effective action $S[J] =- \frac{1}{2}\iint J(x) D(x,y)J(y)$. Here the current is given by $J^{\mu,q} = \frac{it}{2\pi} \epsilon^{\mu \nu \gamma}\partial_\nu A_\gamma^q $. Working in momentum space the full effective action is given by
\begin{equation}
\begin{split}
    S[A] &= \frac{t^2}{16 \pi^2  \tilde{g}} \epsilon^{\mu \alpha \gamma}\epsilon^{\nu \beta \xi} \sum_q\int \frac{d^3k}{(2\pi)^3} k_\alpha A^q_\gamma(k) D^{qq}_{\mu \nu}(k) k_\beta A_\xi(-k)\\
    &=\frac{t^2}{16 \pi^2  \tilde{g}} \sum_q\int \frac{d^3k}{(2\pi)^3} \frac{k^2 (A^q)^2 - \left(k\cdot A^q\right)^2 -i m_q \epsilon^{\mu \nu \gamma}A^q_\mu(k) k_\nu A^q_\gamma(-k)}{k^2-m_q^2} ~~.
    \end{split}
\end{equation}
Restricting attention to the $q=0$ sector, where $m_q = 0$, the Lagrangian has the form 
\begin{equation}
    \mathcal{L}_{q=0} = \frac{t^2}{16 \pi^2  \tilde{g}} A_\mu^{q=0}(k)\left(\eta^{\mu\nu} - \frac{k^\mu k^\nu}{k^2}\right) A^{q=0}_\nu(-k) ~~.
\end{equation}
From this we can surmise that the application of a long wavelength ($q \to 0, k\to 0$ ) vector potential induces a Meissner effect, a hallmark of superconductivity \cite{Zhang1993}.

\subsection{Interaction between vortices}
\label{sec:vortex interaction}
Making the $(z)$ layer index $I$ of the background field manifest, we have seen that vortices are minimally coupled to the gauge field: 
\begin{equation}
    \label{eq:vortex minimal coupling}
	\begin{split}
      \mathcal{L}[a,n]  
      &= \mathcal{L}[a]+\sum_I a^{I}_\mu j^{\mu, I},
     \end{split}
\end{equation}
where $j^{\mu, I}$ is the vortex current in the $I$-th layer and we have used the layered Maxwell-CS theory of section \ref{sec:monopole}:
\begin{equation}
\label{eq:L q basis}
    \mathcal{L}[a] = \frac{K_{IJ}}{4\pi} a_I \wedge da_J +  \tilde{g}\sum_z f_I \wedge \star f_I .
\end{equation}

As before we can diagonalize the K-matrix. For the compressible model i.e. for $m=-(n_1+n_2)$, the normalized eigenvectors $e^q_I = \frac{1}{\sqrt{N}} e^{iqI}$ have eigenvalues $\lambda_q = 2-2\cos q$. In this basis $\mathcal{L} = \sum_q \mathcal{L}_q$ with
\begin{equation}
    \begin{split}
        \mathcal{L}_q = \frac{\lambda_q}{4\pi} b^q \wedge db^{q} +  \tilde{g} f^q \wedge \star f^{q} + b^q \wedge \star j^q.
    \end{split}
\end{equation}

Integrating out the gauge fields, we find the effective action of the currents $j_\mu$ is \cite{ZeeQFT}
\begin{equation}
\label{eq:energy propagator}
E=\frac{1}{2 T} \sum_{q q^\prime}\iint d^{3} x d^{3} y\, j^{\mu}_q(x) D_{\mu \nu}^{q q^\prime}(x, y) j^{\nu}_{q^\prime}(y).
\end{equation} 
where the propagator $D^{\mu\nu}_{qq^\prime}$ is given by Eq. \eqref{eq:propagator}.
Here the measure is given by $d^3x= dx^0 dx^1 dx^2=dtdxdy=dt d^2\vr$. Notice that here $\vr$ denotes the position in the $xy$ plane, different from our convention in the main text.

Consider a vortex-antivortex pair at $(\vr_a, z_a)$ and $(\vr_b,z_b)$ respectively. 
This corresponds to $n(x,y;z) = 2\pi \delta_{z z_a}\delta_{yy_a}\Theta(x-x_a) -2\pi \delta_{z z_b}\delta_{yy_b}\Theta(x-x_b)$ which can be expressed as the following current 
\begin{equation}
j^{\mu}_{(z)} = i\eta^\mu_0[\delta(\vr-\vr_a) \delta_{zz_a} - \delta(\vr-\vr_b) \delta_{zz_b} ] = \sum_q i\eta^\mu_0[\delta(\vr-\vr_a)e^{-iqz_a} - \delta(\vr-\vr_b)e^{-iqz_b}] \frac{e^{iqz}}{\sqrt{N_z}} = \sum_q j^\mu_{q} \frac{e^{iqz}}{\sqrt{N_z}}
\end{equation}
which minimally couples to the gauge field $j^\mu_{(z)} a^{(z)}_\mu = j^{\mu,I} a_\mu^I$.

Plugging this into Eq \ref{eq:energy propagator} and ignoring the self energy terms gives the following
\begin{equation}
    \label{eq:energy calculation}
    \begin{split}
        E &= \frac{1}{2 T} \sum_{q q^\prime}\iint d^{3} x d^{3} y \frac{e^{iqz_a}}{\sqrt{N_z}}\frac{e^{iq^\prime z_b}}{\sqrt{N_z}} D_{qq^\prime}^{00} \delta^2(x-a) \delta^2(y-b)\\
        &=  \frac{-1}{4 T  \tilde{g}} \sum_{q}\iint d x^0 d y^0 \frac{e^{iq(z_a-z_b)}}{N_z}\int \frac{d^3p}{(2\pi)^3} ~ e^{-ix_0p_0}e^{-iy_0p_0} \frac{e^{i\vp\cdot (\vr_a-\vr_b)}}{ p^{2}-\lambda_q^{2}/ \tilde{g}^2} \left(\eta^{00}-\lambda_q^{2} \frac{p^{0} p^{0}}{ \tilde{g}^2p^{4}}\right)\\
        &=  \frac{1}{4 TN_z \tilde{g}} \sum_{q}\int d x^0 e^{iq(z_a-z_b)}\int \frac{dp^1d p^2}{(2\pi)^2} ~ \frac{e^{i\vp\cdot (\vr_a-\vr_b)}}{\vp\cdot\vp + \lambda_q^{2}/ \tilde{g}^2} \\
        &= \frac{1}{2N_z \tilde{g}}\sum_q e^{iq(z_a-z_b)} \int \frac{dp^1d p^2}{(2\pi)^2} ~ \frac{e^{i\vp\cdot (\vr_a-\vr_b)}}{\vp\cdot\vp + \lambda_q^{2}/ \tilde{g}^2}
    \end{split}
\end{equation}
where in going from line 2 to line 3 we first used $\int dx^0 e^{-ix_0p_0} = \delta(p_0)$ and then carried out the integral $\int dp^0$. In going from line 3 to line 4 we used $\int dy^0=T$. In the large $N_z$ limit we can replace $\frac{1}{N_z}\sum_q \to \int dq$, so
\begin{equation}
\begin{split}
    E[r_a,r_b] &\sim \int d^2\vp dq  \frac{e^{i\vp\cdot (\vr_a-\vr_b)}e^{iq(z_a-z_b)}}{\vp^2 + \frac{1}{ \tilde{g}^2}(2-2\cos q)^2}\\
    &\sim \int d^2\vp dq  \frac{e^{i\vp\cdot (\vr_a-\vr_b)}e^{iq(z_a-z_b)}}{\vp^2 + \frac{1}{ \tilde{g}^2}q^4} ~~.
\end{split} 
\end{equation}
To evaluate this we identify $q^2/ \tilde{g}$ with the mass in a 2D free theory. Placing one of the vortices the origin we have
\begin{equation}
\begin{split}
    E[\vr,z] &\sim \int dq e^{iqz } K_0(q^2|\vr|/ \tilde{g}) = \frac{\pi^2}{\sqrt{8}}\frac{\tilde{g}z}{|\vr|}\left[I_{-1/4}\left(\frac{z^2 \tilde{g}}{8|\vr|}\right)^2 - I_{1/4}\left(\frac{z^2 \tilde{g}}{8|\vr|}\right)^2 \right].
    \end{split}
\end{equation}
 Note that in the strong coupling limit $\tilde{g}$ is small.

When $z^2/|\vr| \gg \frac{1}{\tilde{g}}$ then we must use the $I_\nu(x\to\infty)$ asymptotic expansion. 
Note that $K_\nu = \frac{\pi}{2} \frac{I_{-\nu} - I_\nu}{\sin \pi \nu}$, so $I^2_{-\nu}-I^2_{\nu} = \frac{2 \sin \pi\nu}{\pi}K_\nu \left(I_{-\nu}+I_{\nu}\right)$. Combining the asymptotic expansion for $K_{1/4}$ and $I_{\pm1/4}$ gives
\begin{equation}
    E \approx \frac{4\pi}{|z|} + \mathcal{O}\left(\frac{|\vr|^2}{\tilde{g}^2 z^5}\right).
\end{equation}
In the other limit where $z^2/|\vr|$ is comparable or much less than 1 we use the asymptotic form of the limit $I_\nu(x \to 0)$. In this case $I^2_{-1/4}(u)-I^2_{1/4}(u) = \frac{\sqrt{2}}{\Gamma(3/4)^2}\frac{1}{\sqrt{u}} - \frac{\sqrt{u}}{\Gamma(5/4)^2\sqrt{2}} + \hdots$
and so 
\begin{equation}
   \begin{split}
        E \approx \sqrt{\frac{\tilde{g}}{|\vr|}} + \mathcal{O}\left( z^2\frac{\tilde{g}^{3/2}}{|\vr|^{3/2}}\right).
   \end{split}
\end{equation}
Having worked out the basic example of the vortex-antivortex pair we can compute the interaction energy of more complicated configurations of vorticees. Consider the case of a pair of vortex dipoles; one at the origin oriented in the $\hat{z}$ direction and one at position $(\vr_1,z_1)$ oriented in the negative $\hat{z}$ direction. The current is given by 
\begin{equation}
j^{\mu}_{(z)} = i\eta^\mu_0[\delta(\vr-\vr_1) (\delta_{z,z_1} - \delta_{z,z_1+1}) - \delta(\vr) (\delta_{z,0} -\delta_{z,1}) ].
\end{equation}
In the limit where $z_1^2/|\vr_1| \gg \frac{1}{\tilde{g}}$ the interaction energy is given by 
\begin{equation}
    E \approx  \frac{8\pi a_0^2}{|z_1|^3},
\end{equation}
where $a_0$ is the distance between neighboring layers.
In the limit where $z_1^2/|\vr_1|$ is comparable or much less than 1, which includes the case where the dipoles reside in the same $xy$-plane, the interaction energy scales as  \begin{equation}
    E \propto - \left(\frac{\tilde{g}}{|\vr_1|}\right)^{3/2} ~~~.
\end{equation}

\twocolumngrid

\bibliography{CS.bib}

 \end{document}